\documentclass[a4paper,10pt,twoside]{cpc-hepnp}
\usepackage{CJK,upgreek,fancyhdr}
\usepackage{appendix}
\usepackage{revsymb}
\usepackage{multicol}
\usepackage{multirow}
\usepackage{graphicx}
\usepackage{booktabs}
\usepackage{amssymb,bm,mathrsfs,bbm,amscd}
\usepackage[tbtags]{amsmath}
\usepackage{lastpage}
\usepackage[colorlinks=true, urlcolor=blue, linkcolor=red]{hyperref}

\usepackage{xcolor}

\begin{document}
\begin{CJK*}{GB}{gbsn}

\fancyhead[c]{\small Chinese Physics C~~~Vol. xx, No. x (201x) xxxxxx}
\fancyfoot[C]{\small 010201-\thepage}

\footnotetext[0]{Received \today}

\title{\textbf{Measurement of the differential and total cross-sections of $\gamma$-ray emission induced by $14.1$~MeV neutrons for C, Al, Si, Ca, Ti, Cr, and Fe using the tagged neutron method}\thanks{This work was supported by a RSF grant No.23-12-00239}}

\author{%
     Prusachenko P.S.$^{a;1)}$,\email{prusachenko@jinr.ru}%
\quad Grozdanov D.N.$^{a,b}$,
\quad Fedorov N.A.$^{a}$,
\quad Kopatch Yu.N.$^{a}$,
\quad Pampushik G.V.$^{a,c}$,\\
\quad Kharlamov P.I.$^{a,c,d}$,
\quad Skoy V.R.$^{a}$,
\quad Ruskov I.N.$^{a,b}$,
\quad Tretyakova T.Yu.$^{a,c,d}$,
\quad Andreev A.V.$^{a,c}$,\\
\quad Hramco C.$^{a}$,
\quad Filonchik P.G.$^{a,e}$,
\quad and TANGRA collaboration
}

\maketitle

\address{%
$^a${Joint Institute for Nuclear Research (JINR), Dubna, Russia}\\
$^b${Institute for Nuclear Research and Nuclear Energy, Bulgarian Academy of Sciences, Sofia, Bulgaria}\\
$^c${Faculty of Physics, Lomonosov Moscow State University, Moscow, Russia}\\
$^d${Skobeltsyn Institute of Nuclear Physics, Moscow State University, Moscow, Russia}\\
$^e${Moscow Institute of Physics and Technology, Moscow, Russia}
}

\begin{abstract}
\label{abstract}
In this work, differential cross sections of $\gamma$-ray emission produced in nuclear reactions induced by $14.1$~MeV neutrons are measured for the $4.439$~MeV line from carbon, as well as for $10$ individual $\gamma$-ray lines from aluminum, $6$ from silicon, $8$ from calcium, $16$ from titanium, $6$ from chromium, and $14$ from iron. The measurements were conducted using the tagged neutron method with four LaBr$_3$(Ce) scintillation detectors positioned at angles of $25^{\circ}$, $45^{\circ}$, $60^{\circ}$, and $70^{\circ}$ relative to the generator target -- sample center axis. A neutron generator capable of producing $16$ separate beams of tagged neutrons was employed, which, combined with the detector system, enabled the determination of differential cross-sections for $64$ distinct angle values in the range of $17^{\circ}$ to $89^{\circ}$. To simplify data visualization, the angular distributions were divided into $5^{\circ}$ intervals, with weighted mean values of the angle and differential cross-section calculated for each interval. Corrections for multiple neutron scattering and attenuation, $\gamma$-ray attenuation, and total detection efficiency, computed using GEANT4, were accounted for in the cross-section calculations. Additional measurements were performed to validate the correction calculations. The total $\gamma$-ray emission cross-sections were obtained by approximating the angular distributions with even-order Legendre polynomial expansions up to the $6$th degree, followed by integration over the full solid angle. The total systematic error for the obtained data was estimated as $9$\,\%.
\end{abstract}

\begin{keyword}
tagged neutron method, differential and total $\gamma$-ray production cross sections, $14.1$~MeV neutrons
\end{keyword}

\begin{pacs}
25.40, 28.20, 29.25
\end{pacs}

\footnotetext[0]{\hspace*{-3mm}\raisebox{0.3ex}{$\scriptstyle\copyright$}2013
Chinese Physical Society and the Institute of High Energy Physics
of the Chinese Academy of Sciences and the Institute
of Modern Physics of the Chinese Academy of Sciences and IOP Publishing Ltd}%

\begin{multicols}{2}

\section{Introduction}
\label{Introduction}
The study of characteristic $\gamma$-ray emission in nuclear reactions induced by fast neutrons is of significant interest for a range of fundamental and applied problems. In particular, it provides additional information about nuclear structure and the probability of exciting specific nuclear states. Spectroscopy of characteristic $\gamma$-rays emitted in reactions induced by fast neutrons (most commonly with energies around $14$~MeV) is frequently used to investigate the elemental composition of various materials \cite{Karolina2022}. However, it is noted that the accuracy of existing experimental and evaluated data on $\gamma$-ray emission under fast neutron irrays requires substantial improvement to meet modern demands and support emerging fields such as planetary nuclear spectroscopy.

The TANGRA (TAgged Neutrons and Gamma-RAys) project at the Frank Laboratory of Neutron Physics, Joint Institute for Nuclear Research (JINR) \cite{Ruskov2017,Ruskov2021}, employs the tagged neutron method (TNM) \cite{Valkovic2016c2} to address current fundamental and applied challenges \cite{Valkovic2016c7}. The tagged neutron method offers a simple and cost-effective alternative to pulsed neutron generators while retaining many of their advantages (e.g., time-of-flight background suppression, compactness, etc.). Due to these benefits, TNM has been applied to various practical problems, including geology \cite{Bolshakov2020}, metallurgy \cite{Alexakhin2022}, and hazardous material detection \cite{Bishnoi2020}. To enhance the accuracy of analysis and expand its applications, the TANGRA project is conducting a large-scale study of angular distributions and cross-sections for characteristic $\gamma$-ray emission induced by $14.1$~MeV neutrons across a wide range of elements \cite{Grozdanov2018,Fedorov2019,Fedorov2021,Dashkov2022,Kopatch2023,
Kopatch2025}. Among the elements of particular interest are carbon, aluminum, silicon, calcium, titanium, chromium, and iron, which are major constituents of many geological materials and some hazardous substances.

Currently, a considerable amount of experimental data exists for these elements at neutron energies near $14$~MeV, particularly for carbon \cite{Scherrer1953,Benveniste1960,Stewart1964,Morgan1964,
Engesser1967,Clayeux1969,Martin1971,Spaargaren1971,Rogers1975,Bezotosnyi1975,Dickens1977,Zong1979,
Hino1979,Murata1988,Zhou1989,Hasegawa1991,Lashuk1994,Kadenko2016,McEvoy2021,Kelly2021,Kelly2023,
Gordon2025,Morgan1977,HINO1978,Grozdanov2026}, aluminum \cite{Scherrer1953,Bezotosnyi1975,Murata1988,
Lashuk1994,HINO1978,Burymov1969,Bochkarev1965,Pavlik1998,Hlavac1999,Hoot1975,Zhou1997,Nyberg1971,
Hongyu1986,Prud'homme1960}, silicon \cite{Hino1979,Hlavac1999,Prud'homme1960,Boromiza2020,Martin1965,
Abbondanno1973,Connell1975,Bezotosnyj1980,Drosg2002,Zhou2011,Negret2013,Guoying1992}, calcium \cite{Engesser1967}, titanium \cite{Engesser1967,Lashuk1994,Abbondanno1973,Connell1975,Bezotosnyj1980,
Arya1967,Breunlich1971,Morgan1978,Dashdorj2005,Dashdorj2007,Olacel2017}, chromium \cite{Burymov1969,
Abbondanno1973,Arya1967,Breunlich1971,Kinney1972,Grenier1974,Voss1975,Yamamoto1978,Oblozinsky1992,
Mihailescu2007}, and iron \cite{Zong1979,Hasegawa1991,Hongyu1986,Prud'homme1960,Abbondanno1973,
Bezotosnyj1980,Arya1967,Kinney1972,Yamamoto1978,Bostrom1959,Joensson1969,Sukhanov1970,Degtyarev1977,
Drake1978,Xiamin1982,Shalabi1983,Hlavac1983,Jinqiang1988,Sakane1999,Negret2014,Broder1970,Western1965,
Mitsuda2002,Antalik1980,Nelson2005}. Despite the impressive number of studies, one of the major issues is the fragmented and insufficient data on angular distributions of $\gamma$-rays emitted in neutron-induced reactions. According to Simakov's compilation \cite{Simakov1998}, which provides a detailed review of experimental data available up to 1998 (see Table 3 in \cite{Simakov1998}), most measurements were performed for only $1-2$ angular points, making it impossible to accurately assess $\gamma$-ray emission anisotropy or reliably estimate total cross-sections. Additionally, many studies lack detailed descriptions of experimental setups and data correction methodologies, complicating efforts to resolve discrepancies in reported cross-sections. Some recent studies, such as those conducted at the GAINS setup in GELINA \cite{Negret2013,Olacel2017,Negret2014}, provide thorough descriptions of experimental procedures and data analysis. However, their limitations include a narrow selection of studied elements and the absence of angular distribution measurements.

The objective of this research was to perform more precise and detailed measurements of differential $\gamma$-ray emission cross-sections for the most intense transitions in carbon, aluminum, silicon, calcium, titanium, chromium, and iron nuclei under irrays by $14.1$~MeV neutrons, followed by the determination of total cross-sections. The detector and sample geometry, as well as sample dimensions, were optimized to minimize systematic uncertainties, such as multiple scattering, $\gamma$-ray attenuation, and incomplete sample coverage by tagged neutron beams.

\section{Experiment}
\label{Experiment}
The tagged neutron method is based on the detection of secondary $\alpha$-particles accompanying neutron emission in the $^3$H$(d,n)^4$He fusion reaction using a position-sensitive charged particle detector integrated into the neutron generator vacuum chamber. This provides information about the neutron emission time, direction, and its flux. A characteristic feature of this method is the relatively large size of the tagged neutron beam field, which necessitates either using large samples or accepting that not all neutrons in the beam will hit the sample surface. This can introduce significant systematic effects that distort measurement results, including uncertainty in determining the number of neutrons incident on the sample, as well as absorption and multiple scattering of both $\gamma$-rays and neutrons in the sample. In this work, the geometric parameters of the setup were optimized to minimize these effects. In particular, the sample size was selected to ensure complete interception of all tagged neutron beams, while the thickness was reduced compared to previous experiments in the TANGRA project. A series of additional experiments were conducted to evaluate the accuracy of correction factors and measurement uncertainty limits.

The general layout of the experimental setup is shown in Fig.~\ref{fig:fig1}. The neutron source was an ING-27 generator \cite{Prusachenko2025} with accelerated deuteron energies of $30$ -- $90$~keV. The $\alpha$-particles accompanying neutron emission were detected using a built-in position-sensitive charged particle detector. This detector consists of 16 vertical and 16 horizontal strips forming 256 pixels. Each pixel size was $4\times 4$~mm$^2$, and the distance between the tritium target and $\alpha$-particle detector was $44$~mm. In this work, to increase counting statistics, data from only 16 vertical strips were used without pixel subdivision. A more detailed description of the neutron generator used is provided in \cite{Capote2009}.

The experiment utilized samples of graphite, silicon and chromium oxides, as well as metallic aluminum, titanium and iron with natural isotopic abundance. Each sample measured $44\times44$~cm, a size selected to ensure complete interception of all tagged neutron beams at the chosen target-to-sample distance of $24.8$~cm. The mass of each sample was precisely measured using precision electronic scales. Sample thickness ranged from $0.7$ to $2$~cm. Powdered materials (SiO$_2$, CaO and Cr$_2$O$_3$) were contained in specially fabricated thin polyethylene boxes ($3$~mm wall thickness). Detailed specifications for each sample used in the study are provided in Table~\ref{table:samples}.

\includegraphics[width=75mm]{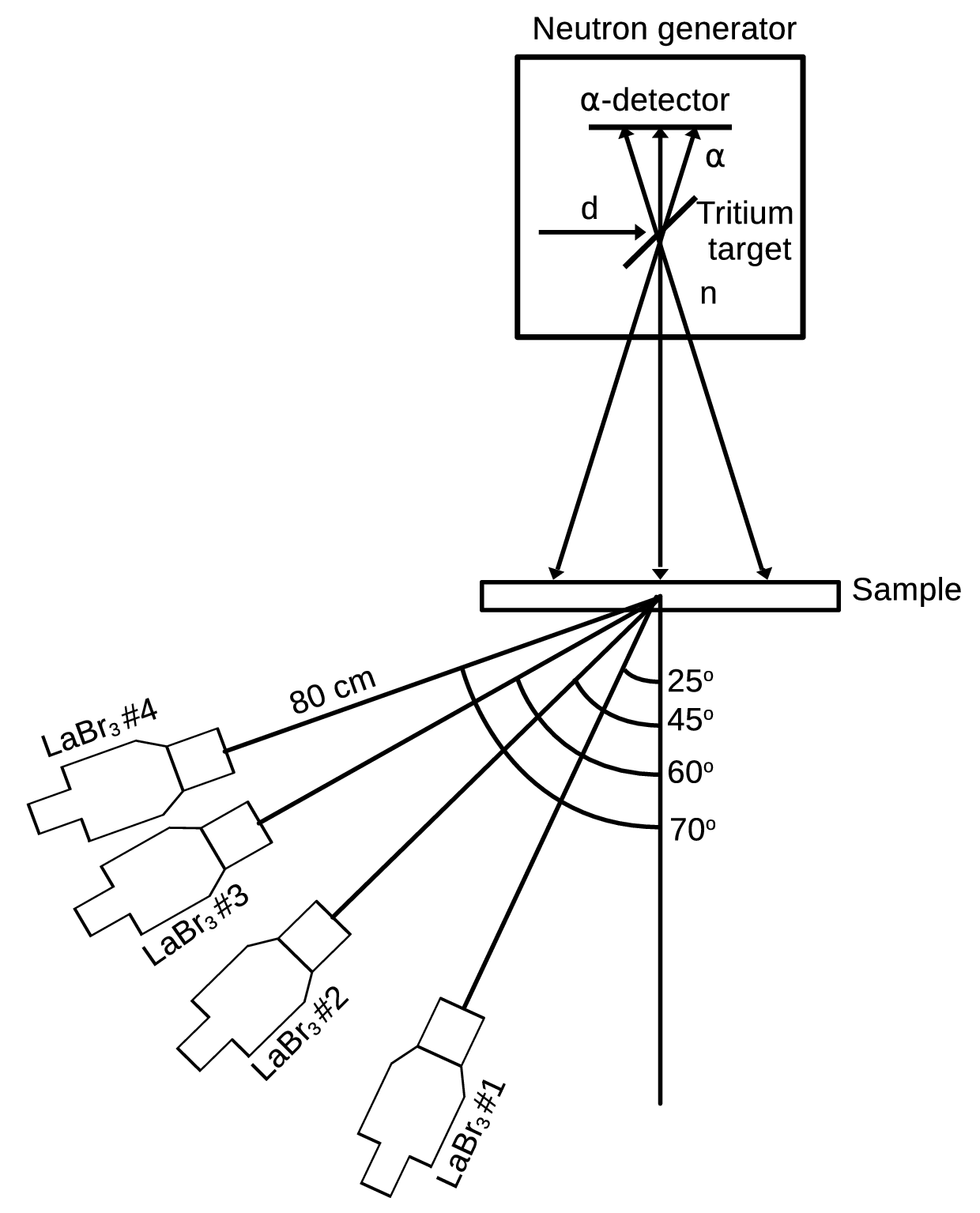}
\figcaption{Layout of the experimental setup (not to scale). $d$, $n$, $\alpha$ -- designations of deuteron beams, tagged neutrons and $\alpha$-particles, respectively.}
\label{fig:fig1}

The experiment utilized samples of graphite, silicon and chromium oxides, as well as metallic aluminum, titanium and iron with natural isotopic abundance. Each sample measured $44\times44$~cm, a size selected to ensure complete interception of all tagged neutron beams at the chosen target-to-sample distance of $24.8$~cm. The mass of each sample was precisely measured using precision electronic scales. Sample thickness ranged from $0.7$ to $2$~cm. Powdered materials (SiO$_2$, CaO and Cr$_2$O$_3$) were contained in specially fabricated thin polyethylene boxes ($3$~mm wall thickness). Detailed specifications for each sample used in the study are provided in Table~\ref{table:samples}.

The detector system consisted of $4$ LaBr$_3$(Ce) scintillation detectors with crystal dimensions of $3\,'' \times 3\,''$. The detectors were positioned at equal distances ($80$~cm) from the sample center in the horizontal plane at angles of $25^{\circ}$, $45^{\circ}$, $60^{\circ}$, and $70^{\circ}$ relative to the axis connecting the generator target and the sample center. The combination of $16$ vertical $\alpha$-detector strips and $4$ $\gamma$-detectors allowed obtaining $64$ $\gamma$-ray emission angles in the range from $17^{\circ}$ to $89^{\circ}$. The weighted average $\gamma$-ray emission angle and neutron incidence angle were determined by Monte Carlo simulations for each detector-strip combination.

The experimental procedure included separate measurements with each sample, as well as measurements without a sample and with an empty container (for SiO$_2$, CaO and Cr$_2$O$_3$ samples) to account for time-of-flight dependent background. The core of the data acquisition system was a $128$-channel waveform digitizer with a sampling rate of $100$~MSamples/sec and $16$-bit analog-to-digital converter (ADC) resolution. Signals from each $\alpha$-detector strip and each scintillation detector were digitized. At the digital signal processing stage, the main characteristics of each signal were determined, including its time-stamp, amplitude, and pulse area, which is proportional to the light output in the case of scintillation detectors. Then, from the entire event array, coincidences between signals from the $\alpha$-detector and $\gamma$-ray detectors were selected, and subsequently time and amplitude distributions were constructed for each detector-strip combination. The time-of-flight was estimated as the time difference between signals from the $\gamma$-detector and $\alpha$-detector. The obtained spectra served as the basis for subsequent processing.

\end{multicols}
\begin{center}
\tabcaption{\label{table:samples} Specifications of the samples used in the current work.}
\footnotesize
\begin{tabular}{ c c c c c }
\hline
\textbf{Sample} & \textbf{Dimensions (cm)} & \textbf{Purity (\%)} & \textbf{Mass (g)} & \textbf{Density (g/cm$^3$)} \\ \hline \hline
Graphite (C) & $44\times44\times2$    & 99                 & 6670  & 1.64 \\ \hline
Al       & $44\times44\times0.76$ & 99                 & 3978  & 2.70 \\ \hline
SiO$_2$     & $44\times44\times2$    & 99                 & 2418  & 0.62 \\ \hline
Ti       & $44\times44\times0.9$  & \textgreater{}99.5 & 7601  & 4.36 \\ \hline
Cr$_2$O$_3$    & $44\times44\times2$    & \textgreater{}99   & 5161  & 1.33 \\ \hline
Fe       & $44\times44\times0.9$  & 97                 & 13614 & 7.81 \\ \hline
CaO      & $44\times44\times2$    & 99                 & 2335  & 0.60 \\ \hline
\end{tabular}
\end{center}
\begin{multicols}{2}

\section{Data analysis}
\label{Data analysis}

\subsection{Analysis of spectra}
\label{Analysis of spectra}

Examples of one-dimensional time-of-flight (TOF) distributions for measurements with a titanium sample and without a sample are shown in Fig.~\ref{fig:fig2}. As can be seen from the figure, the spectrum corresponding to the measurement with the sample shows three groups of events, which can be separated by TOF. The first group corresponds to the emission of prompt $\gamma$-rays from reactions induced by fast neutrons in the tritium target surroundings (substrate and generator housing), the second group corresponds to the detection of $\gamma$-rays resulting from reactions in the sample, and the third group corresponds to neutrons scattered in the sample and generator that hit the gamma detector. In the case of measurements without a sample, only two groups of events are observed, corresponding to prompt $\gamma$-rays from the generator and scattered neutrons.

\vspace{12pt}
\includegraphics[width=75mm]{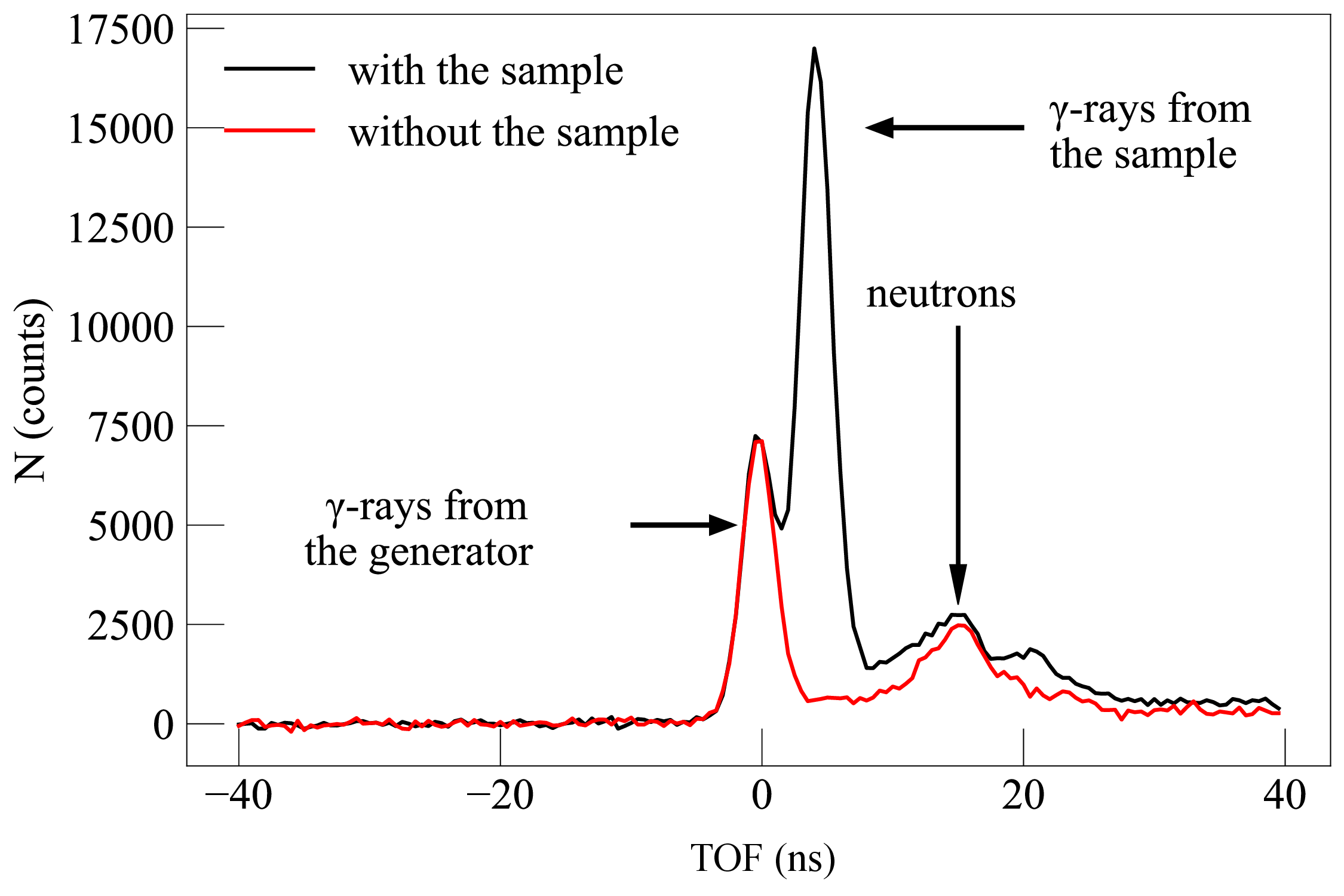}
\figcaption{An example of one-dimensional TOF distributions for measurements with and without a titanium sample, corresponding to the combination "first strip -- first detector" (scattering angle $54^{\circ}$).}
\label{fig:fig2}

\end{multicols}

\includegraphics[width=160mm]{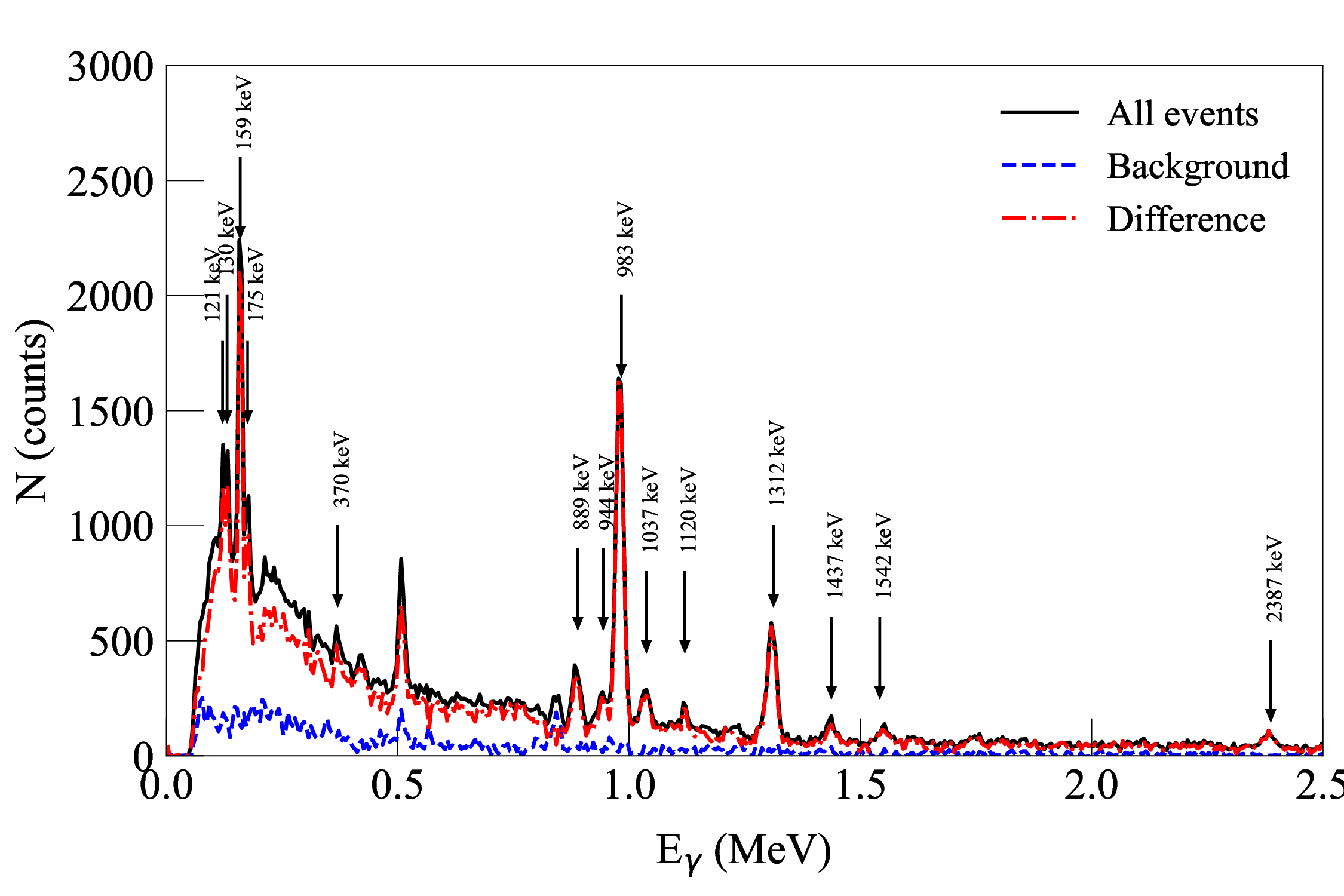}
\figcaption{An example of amplitude spectra before and after background subtraction, corresponding to the combination "first strip -- first detector" (scattering angle $54^{\circ}$). The presented spectra correspond to measurements with the titanium sample, without the sample, and their difference, constructed for the time window of $\pm3\sigma$ from the sample peak position on the TOF scale. The amplitude scale is calibrated in units of $\gamma$-ray energy. The $\gamma$-ray energies shown in the figure correspond to the experimentally observed $\gamma$-transitions in titanium nuclei.}
\label{fig:fig3}

\begin{multicols}{2}

At the first stage of spectrum analysis, the random coincidence background was subtracted for all detector-strip combinations. For this purpose, an amplitude spectrum was built in the TOF window corresponding to $\pm3\sigma$ from the sample peak position, from which the random coincidence amplitude spectrum constructed in the window from $-250$~ns to $-50$~ns was subtracted, taking into account the time window width. This procedure was performed both for spectra obtained from measurements with and without the sample. At the next stage, one-dimensional amplitude spectra were constructed for the same time windows for measurements with the sample and corresponding measurements without the sample. Examples of such spectra are shown in Fig.~\ref{fig:fig3}. The background amplitude spectra were subtracted from the spectra measured with the sample, taking into account the difference in the number of registered $\alpha$-particles and the effect of shielding the $\gamma$-background from the generator by the sample. The latter was estimated by GEANT4 simulations. Thus, spectra corresponding to the registration of only $\gamma$-rays from the sample were obtained.

After that, the spectra were decoded, full-energy absorption peaks corresponding to the expected transitions in the studied nuclei were identified according to information from the RIPL-3 database \cite{Capote2009}, which is practically identical to ENSDF \cite{ENSDF}. The areas of these peaks were determined from a Gaussian function fit with a linear substrate.

\subsection{Determination of differential and total cross-sections for individual $\gamma$-lines}
\label{Determination of differential and total cross-sections for individual}
The differential cross-section of $\gamma$-ray emission was calculated according to the following expressions:

\begin{equation}
\label{eq:1}
\frac{d\sigma}{d\Omega}(\theta) = \frac{N_p(\theta\cos\xi)}{4 \pi N_{\alpha} n_{nucl} k} \cdot 10^{27} \left[\frac{\mbox{mb}}{\mbox{sr}}\right],
\end{equation}

\begin{equation}
\label{eq:2}
k = k_{na} \int\limits_{0}^{x_0} \epsilon(x) k_{ms}(x) k_{\gamma\alpha}(x) dx,
\end{equation}
where $N_p$ is the full-energy peak area corresponding to the current detector-strip combination; $N_{\alpha}$ is a number of registered $\alpha$-particles from the $^3$H$(d,n)^4$He reaction for the current strip, corresponding to the number of emitted tagged neutrons; $n_{nucl}$ is a surface density of nuclei that induced reactions, which lead to formation of the $\gamma$-peak; $\xi$ is an average angle of neutron incidence on the sample for the current strip; $k$ is an integral correction accounting for attenuation of the tagged neutron beam in the neutron generator $k_{na}$, total detection efficiency $\epsilon$, contribution of $\gamma$-rays resulting from multiple neutron scattering in the sample $k_{ms}$, and absorption or energy change of $\gamma$-rays due to interactions in the sample$k_{\gamma\alpha}$; $x_0$ is the sample thickness. Expression (\ref{eq:2}) allows proper consideration of changes in total detection efficiency and other corrections depending on the depth $x$ at which the interaction occurred in the sample. It should be noted that the cross sections obtained in our work are given for $\gamma$-lines attributed to a specific isotope or, in cases where the observed peak contains unresolved lines from several isotopes, for the sum of these isotopes. Accordingly, for each observed line, the surface density of nuclei $n_{nucl}$ in formula (\ref{eq:1}) was calculated taking into account the abundance of all isotopes of each element whose reactions could contribute to the formation of the studied photopeak:

\begin{equation}
\label{eq:3}
n_{nucl} = \sum\limits_{0}^{j} a_i c_i,
\end{equation}
where $c$ is the concentration of a specific isotope in the natural mixture on which the reaction occurs producing the target $\gamma$-ray line, $a_i$ is the coefficient representing amount of considered element in the empirical formula of the substance and $j$ is a number of reaction channels that may contribute to formation of the discussed photopeak.

The correction for absorption of primary neutrons in the target substrate ($2$~mm copper, tilt angle $45^{\circ}$) and the neutron generator wall ($1.5$~mm steel) was calculated in a separate GEANT4 \cite{Agostinelli2003,Allison2016} simulation for each strip. The calculation results showed that the number of neutrons reaching the sample decreases by $9$\,\% compared to the number emitted, due to absorption and large-angle scattering. Meanwhile, approximately $1.5-2$\,\% of the total neutrons reaching the sample have energies below $14$~MeV. However, additional calculations demonstrated their negligible contribution to the total yield of emitted $\gamma$-rays.

The product of the total $\gamma$-ray detection efficiency $\epsilon$ and the $\gamma$-ray attenuation coefficient $k_{\gamma a}$ as a function of depth in the sample was calculated using the Monte Carlo method in GEANT4 individually for each strip-detector combination. It was determined as the ratio of emitted $\gamma$-rays to registered full-energy peak events. A distinctive feature of this procedure was that $\gamma$-rays were emitted from the sample region corresponding to a specific strip, while accounting for the decreasing probability of $\gamma$-ray emission with increasing depth due to absorption of primary tagged neutrons. The calculations were verified in several dedicated experiments. In the low-energy range ($0.1-3$~MeV), the total detection efficiency was measured using standard $\gamma$-ray isotopic sources ($^{22}$Na, $^{60}$Co, $^{137}$Cs, $^{133}$Ba, $^{228}$Th) with known activity ($4$\,\% uncertainty). The sources were positioned at the center of the sample's front plane.

A second experiment measured the relative detection efficiency for high-energy $\gamma$-rays. In this setup, a $5$-liter container filled with concentrated NaCl solution was placed at the sample position. A $^{239}$PuBe neutron source was inserted to the center of this container. Some neutrons from the source were thermalized and subsequently captured by chlorine nuclei in the solution. Analysis of the measured spectra identified the most intense $\gamma$-lines above $3$~MeV, corresponding to transitions in $^{36}$Cl nuclei at $7.413$, $7.79$, and $8.578$ MeV (energy values from the prompt $\gamma$-ray database \cite{Richard2003}). Key selection criteria required these lines to be free from single/double escape peak interference from higher-energy $\gamma$-rays. Contributions from $^{16}$O$(n,\gamma)^{17}$O and $^{37}$Cl$(n,\gamma)^{38}$Cl reactions were considered negligible due to extremely small cross-sections. The $1.951$ and $1.959$~MeV lines served as reference transitions in the $0.1-3$~MeV range where efficiency could be verified using isotopic sources. The relative detection efficiency was then determined using the equation:
\begin{equation}
\label{eq:4}
\epsilon_{\gamma}(E) = \frac{N_p(E)k(E)Y(E)}{N_p(1.951~\mbox{MeV})k(1.951~\mbox{MeV})Y(1.951~\mbox{MeV})},
\end{equation}
where $N_p$ is the area of the full-energy peak for the corresponding $\gamma$-ray energy, $k$ is a correction accounting for $\gamma$-ray absorption in the source volume, and $Y$ is a yield for a specific $\gamma$-ray line. The yield values and their uncertainties were also taken from the IAEA database \cite{Richard2003}. The efficiency values obtained in this way were added to the results of measurements with isotopic sources after normalization. 

\vspace{12pt}
\includegraphics[width=75mm]{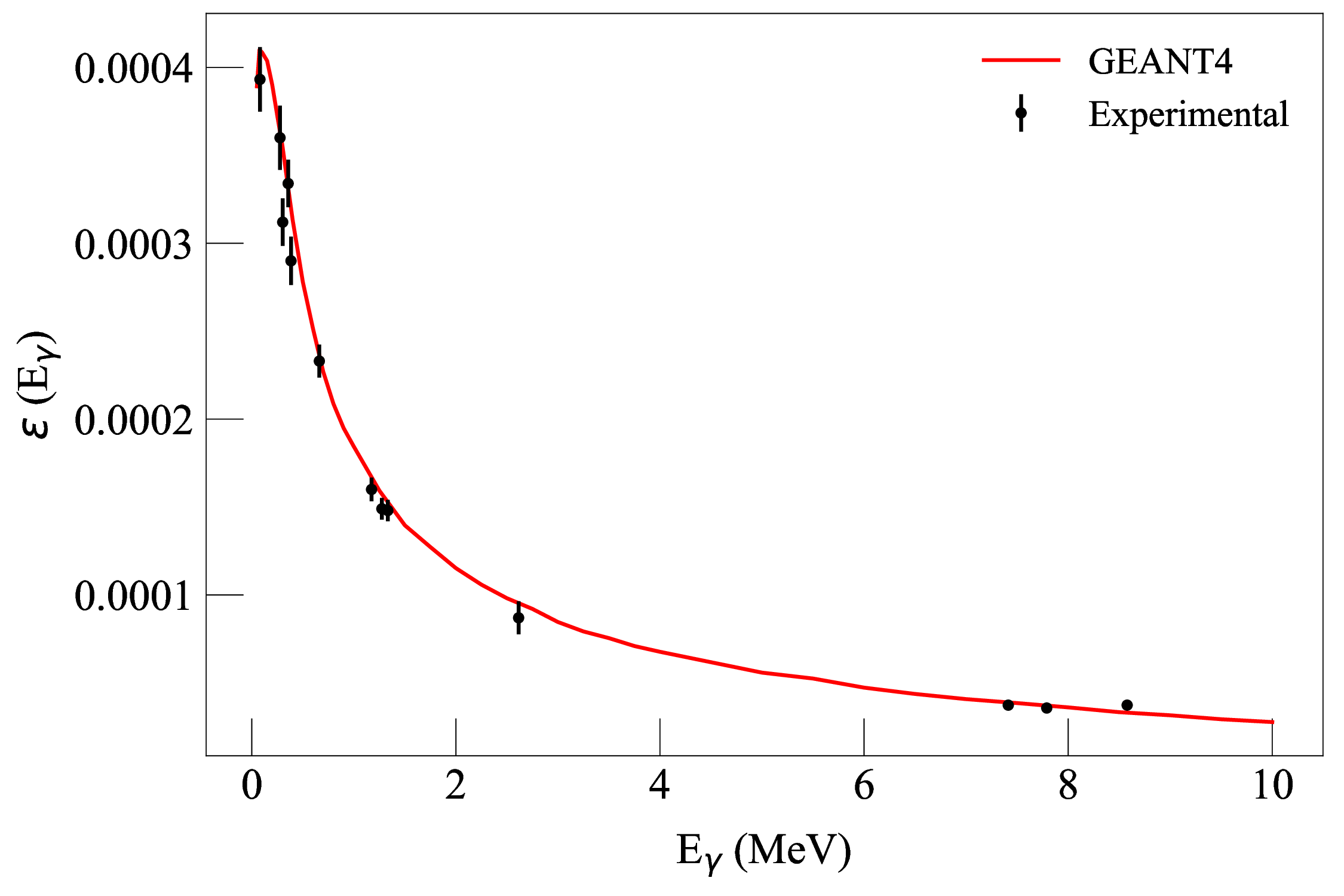}
\figcaption{Comparison of experimental and calculated total detection efficiency for one of the LaBr$_3$(Ce) detectors.}
\label{fig:fig4}

The results of comparing the simulated and experimental efficiency for one of the detectors are shown in Fig.~\ref{fig:fig4}. As can be seen from the figure, the simulation results generally agree with the experimental data within the measurement uncertainties. The average deviation between experimental points and the model curve was about $5$\,\%. This value was subsequently used as an estimate of the efficiency uncertainty. To verify the accuracy of the attenuation coefficient calculation, a series of additional measurements was performed with $^{137}$Cs and $^{60}$Co sources. In these measurements, the sources were placed at the center of iron and titanium samples on the side opposite to the detectors. A measurement was also performed with a dummy sample made of thin polystyrene foam, placed in such a way as to maintain the geometry of the detector and source arrangement. The attenuation coefficient for a specific $\gamma$-ray line was calculated as the ratio of the full-energy peak area with the sample to the corresponding peak area without the sample. Comparison of the obtained results with GEANT4 simulation results showed that the difference between experiment and calculation does not exceed $2$\,\% in all cases. This value was used as the maximum estimate of the uncertainty associated with the calculation of the absorption coefficient.

Separately, a simulation of the coefficient $k_{ms}$ was performed, which accounted for multiple neutron scattering in the sample with subsequent generation of secondary $\gamma$-rays. Direct calculation of this correction in GEANT4 is extremely difficult, primarily due to the inability to account for the influence of $(n,2n)$ reactions, since GEANT4 does not model the residual nucleus and its de-excitation for this reaction channel. To partially solve this problem, a two-stage calculation was performed. At the first stage, GEANT4 simulated the transport of neutrons emitted from the generator target through the sample. Here, the sample model was divided into thin layers, and neutron spectra were calculated for each layer. In general, the correction $k_{ms}$ for $\gamma$-rays with the required energy was calculated as the ratio of the calculated number of $\gamma$-rays generated by all neutrons $N_{\gamma}^{tot}$ to the number generated only by primary neutrons $N_{\gamma}^{i}$, according to the following expression:
\begin{equation}
\label{eq:5}
k_{ms}(x) = \frac{N_{\gamma}^{tot}(x)}{N_{\gamma}^{i}(x)} = \frac{ \sum\limits_{0}^{j} \int\limits_{0}^{14.1} F(E,x) n_j \sigma_j(E)dE}{\sum\limits_{0}^{j} F(14.1~\mbox{MeV},x) n_j \sigma_j (14.1~\mbox{MeV}))},
\end{equation}
where $x$ is the depth in the sample; $j$ is the number of reaction channels leading to the emission of $\gamma$-rays with the required energy; $F(E,x)$ is the number of neutrons with energy $E$ at depth $x$; $\sigma_j(E)$ is the cross-section of $\gamma$-ray emission with the required energy induced by neutrons with energy $E$ for the $j$-th reaction channel; $n_j$ is the surface density of atoms of the isotope on which the $j$-th reaction occurs. The energy dependences of the emission cross-sections for the $\gamma$-ray energies observed in the experiment were obtained for all possible reaction channels using the TALYS-2.1 code \cite{Koning2023} with default parameters. TALYS was chosen because it calculates partial cross-sections for residual nucleus formation in $(n,2n)$ reactions, unlike other standard libraries such as ENDF/B-VIII.0 \cite{Brown2018} or JENDL-5 \cite{Iwamoto2023}. For carbon, no significant contribution from $(n,2n)$ reactions was expected due to the low content of isotopes other than $^{12}$C, so the $^{12}$C$(n,n')^{12}$C reaction cross-section from ENDF/B-VIII.0 was used in this case. The calculation results showed that the contribution of multiple scattering ranges from $5-10$\,\%, depending on the depth in the sample, for lines corresponding to transitions from low-lying states (e.g., $983$~keV for titanium or $846$~keV for iron) to $15-25$\,\% for some individual lines characterized by a very strong dependence of the cross-section on neutron energy, according to TALYS estimates.

To estimate the uncertainty of the correction factor related to the accuracy of the cross-sections provided by TALYS and ENDF, a separate series of calculations was performed using available experimental data on the energy dependence of emission cross-sections for some $\gamma$-lines of C, Si, Ti, and Fe, previously measured at LANL \cite{Kelly2023} and GELINA \cite{Martin1965,Negret2013,
Negret2014}. The difference between the cross-sections obtained using TALYS and experimental data was used to estimate the method's uncertainty. Depending on the specific line and element, the contribution of this error was $3-7$\,\%, with $7$\,\% adopted as the upper uncertainty estimate for all lines.

To verify the accuracy of the correction calculations, additional measurements of $\gamma$-ray emission cross-sections were performed for iron samples of different thicknesses ($3$~mm and $18$~mm), in addition to the $9$~mm sample used in the main measurements. Among all the samples used in the measurements, iron has the highest effective $Z$ and density, leading to the largest expected influence of neutron multiple scattering and $\gamma$-ray attenuation. The maximum difference between the cross-sections of the same $\gamma$-lines obtained for the thinnest and thickest samples did not exceed $7$\,\%, which is within the estimated systematic error of this experiment ($9$\,\%, see Section \ref{Measurement uncertainties}).

To obtain the total cross-section, the corresponding differential cross-sections calculated using formula (\ref{eq:1}) were approximated in form of a Legendre-polynomial expansion of even order:
\begin{equation}
\label{eq:6}
\frac{d\sigma}{d\Omega}(\theta) = \frac{\sigma_{\gamma}}{4\pi} \sum\limits_{\nu=0}^{2J} a_{\nu} P_{\nu} (\cos\theta).
\end{equation}
%In this work, only even-order Legendre polynomials were used to describe the angular distributions of $\gamma$-ray emission. Furthermore, in the representation given by
In this expression the coefficient $a_0$ is equal to $1$, and $J$ is a multipole of the considered $\gamma$-transition. 

\subsection{Measurement uncertainties}
\label{Measurement uncertainties}

The main sources of systematic uncertainty in this experiment are given in Table~\ref{table:Systematic}. The total efficiency uncertainty was estimated as the average difference between efficiency values calculated in GEANT4 and experimental data. The upper limit of uncertainty associated with attenuation of secondary $\gamma$-rays in the sample was determined in a separate experiment with $^{137}$Cs and $^{60}$Co sources (see section \ref{Determination of differential and total cross-sections for individual}). The largest contribution to the systematic uncertainty of this experiment came from the multiple scattering correction in the sample. Its upper limit was estimated by comparing correction coefficients obtained from estimated cross-sections from TALYS code for heavy nuclei (heavier than oxygen), ENDF/B-VIII.0 library for carbon, and available experimental data from literature on energy dependence of emission of individual $\gamma$-lines. The uncertainty in the number of nuclei in the sample included both the uncertainty in its mass determination and the uncertainty related to purity of some samples. The total systematic uncertainty was estimated as $9$\,\%. The statistical uncertainty for differential cross-sections varied from $0.5-1$\,\% for the most intense lines to $20-30$\,\% for the least intense lines.

\begin{center}
\tabcaption{\label{table:Systematic} Systematic uncertainty budget in the experiment.}
\footnotesize
\begin{tabular}{ p{5cm} c }
\hline
Source                                                  & Contribution (\%) \\ \hline \hline
Efficiency                                              & 5               \\ \hline
Correction for attenuation of $\gamma$-rays in the sample & 2               \\ \hline
Multiple neutron scattering                             & 7               \\ \hline
Number of nuclei in the sample                          & 2               \\ \hline
Total                                                   & 9               \\ \hline
\end{tabular}
\end{center}

\newpage
\section{Results and Discussion}
\label{Results and Discussion}

\subsection{Carbon}
\label{Carbon}

The data on $\gamma$-ray emission from the carbon sample are represented solely by the $4.439$~MeV line (see  Fig.~\ref{fig:fig5} and Table~\ref{table:C}), corresponding to the $2^+ \longrightarrow 0^+$ transition in the $^{12}$C nucleus excited via the $^{12}$C$(n,n')^{12}$C reaction. The contribution from the $^{13}$C$(n,2n)^{12}$C reaction is assumed to be negligible ($<1$\,\%). As shown in Fig.~\ref{fig:fig5} (a), the angular distribution of $4.439$~MeV $\gamma$-rays measured in the present work agrees within uncertainties with the data reported in Ref.~\cite{Kelly2021} for angles greater than $90^{\circ}$, as well as with the data from Refs.~\cite{Engesser1967,Clayeux1969,Martin1971,
Zong1979,Zhou1989,Hasegawa1991,McEvoy2021}, which provide measurements only for angles close to $90^{\circ}$. The differential cross-sections presented in Refs.~\cite{Benveniste1960,Spaargaren1971}, as well as in Ref.~\cite{Kelly2021} for the angular range of $0-90^{\circ}$, significantly exceed the data obtained in this work, with Ref.~\cite{Kelly2021} exhibiting a notable asymmetry in the angular distribution. Additionally, all experimental angular distributions reported over a wide angular range predict a significant $a_4$ coefficient in the Legendre polynomial expansion, whereas the evaluated data from ENDF/B-VIII.0 include only the $a_2$ coefficient.

As shown in Fig.~\ref{fig:fig5} (b), there is also some discrepancy in the experimental data on total emission cross-sections. Specifically, the total cross-section obtained in this work is noticeably smaller ($\approx17$\,\%) than the data from direct $\gamma$-ray measurements in Ref.~\cite{Kelly2021} ($224$~mb), ($\approx10$\,\%) larger than the cross-section from Ref.~\cite{Rogers1975} ($173$~mb), and agrees within uncertainties with the cross-sections derived from the $n-\gamma$ correlation experiment in \cite{Kelly2023}, as well as those reported in Ref.~\cite{Gordon2025} and the evaluated cross-section from the ENDF/B-VIII.0 library \cite{Brown2018} ($210$~mb). The cross-section values obtained in a recent experiment within the TANGRA project \cite{Grozdanov2026} using an array of plastic scintillation detectors are also consistent with the results presented here, both for the case where the cross-section was determined from the $\gamma$-ray angular distribution ($205\pm11$~mb) and for the case where neutrons corresponding to the same scattering channel were detected ($192\pm10$~mb). However, there is a slight difference in the Legendre polynomial expansion coefficients, though it remains within the uncertainties reported in \cite{Grozdanov2026}. It is worth noting separately that the results presented in \cite{Kelly2023}, both for the direct measurement of the $\gamma$-ray emission cross-section and for the $n-\gamma$ correlation experiment, represent relative measurements of the energy dependence of the cross-section, normalized to the ENDF/B-VIII.0 evaluated cross-section. 

\vspace{12pt}
\includegraphics[width=75mm]{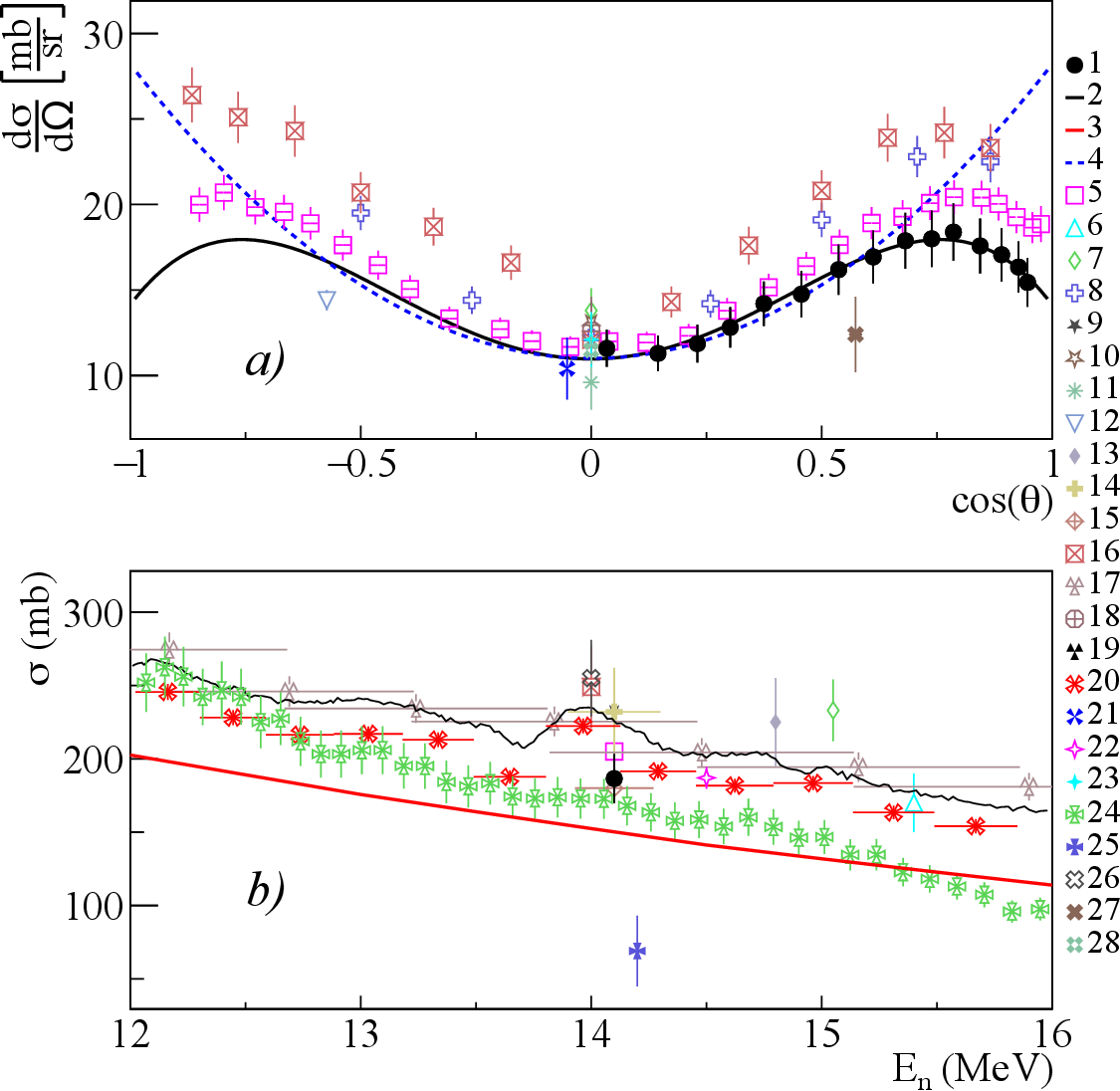}
\figcaption{Differential \textbf{(a) and total (b)} cross sections of $\gamma$-ray emission with energy $4.439$~MeV from the $^{12}$C$(n,n')^{12}$C reaction in comparison with experimental data from other authors, evaluated cross sections from ENDF/B-VIII.0 libraries, and theoretical calculations based on the TALYS program with default parameters. 1 -- data from present work; 2 -- angular distribution approximation from present work using Legendre polynomials; 3 -- TALYS calculation; 4 -- evaluated cross section from ENDF/B-VIII.0 \cite{Brown2018}; 5 -- \cite{Grozdanov2026}; 6 -- \cite{Lashuk1994}; 7 -- \cite{McEvoy2021}; 8 -- \cite{Spaargaren1971}; 9 -- \cite{Stewart1964}; 10 -- \cite{Engesser1967}; 11 -- \cite{Clayeux1969}; 12 -- \cite{Morgan1977}; 13 -- \cite{Morgan1964}; 14 -- \cite{Kadenko2016}; 15 -- \cite{Murata1988}; 16 -- \cite{Benveniste1960}; 17 -- \cite{Gordon2025}; 18 -- \cite{Hasegawa1991}; 19, 20 -- \cite{Kelly2023}, direct $\gamma$-ray measurement and correlated $n-\gamma$ measurements; 21 -- \cite{Zong1979}; 22 -- \cite{Simakov1998}; 23 -- \cite{Martin1971}; 24 -- \cite{Rogers1975}; 25 -- \cite{Scherrer1953}; 26 -- \cite{Bezotosnyi1975}; 27 -- \cite{HINO1978}; 28 -- \cite{Zhou1989}.}
\label{fig:fig5}

\end{multicols}

\begin{center}
\tabcaption{\label{table:C} Total emission cross-section $\sigma$ and angular distribution decomposition coefficients  into Legendre polynomials $a_2$ and $a_4$ for the $4.439$~MeV $\gamma$-ray line emitted in the interaction of $14.1$~MeV neutrons with carbon nuclei. The energies of the initial ($i$) and final ($f$) states are given in MeV.}
\footnotesize
\begin{tabular}{ c c c c c c }
\hline
$E_{\gamma}$~(MeV) & Reaction     & Transition, $E_{i}$ $(J_{i}^{\pi}) \rightarrow E_{f} (J_{f}^{\pi})$ & $\sigma$~(mb) & $a_2$      & $a_4$         \\ \hline
 \hline
$4.439$   & $^{12}$C$(n,n')^{12}$C & $4.439(2^+) \rightarrow $g.s.$(0^+)$ & $186\pm17$ & $0.28\pm0.01$ & $-0.33\pm0.02$ \\ \hline
\end{tabular}
\end{center}

\begin{multicols}{2}

\subsection{Aluminum and Silicon}
\label{Aluminum and silicon}
The measurement results for the most intense $\gamma$-ray lines emitted during the interaction of $14.1$~MeV neutrons with silicon and aluminum nuclei are presented in Figs.~\ref{fig:fig6} -- \ref{fig:fig9} and Table~\ref{table:Al} and \ref{table:Si}.

\vspace{12pt}
\includegraphics[width=75mm]{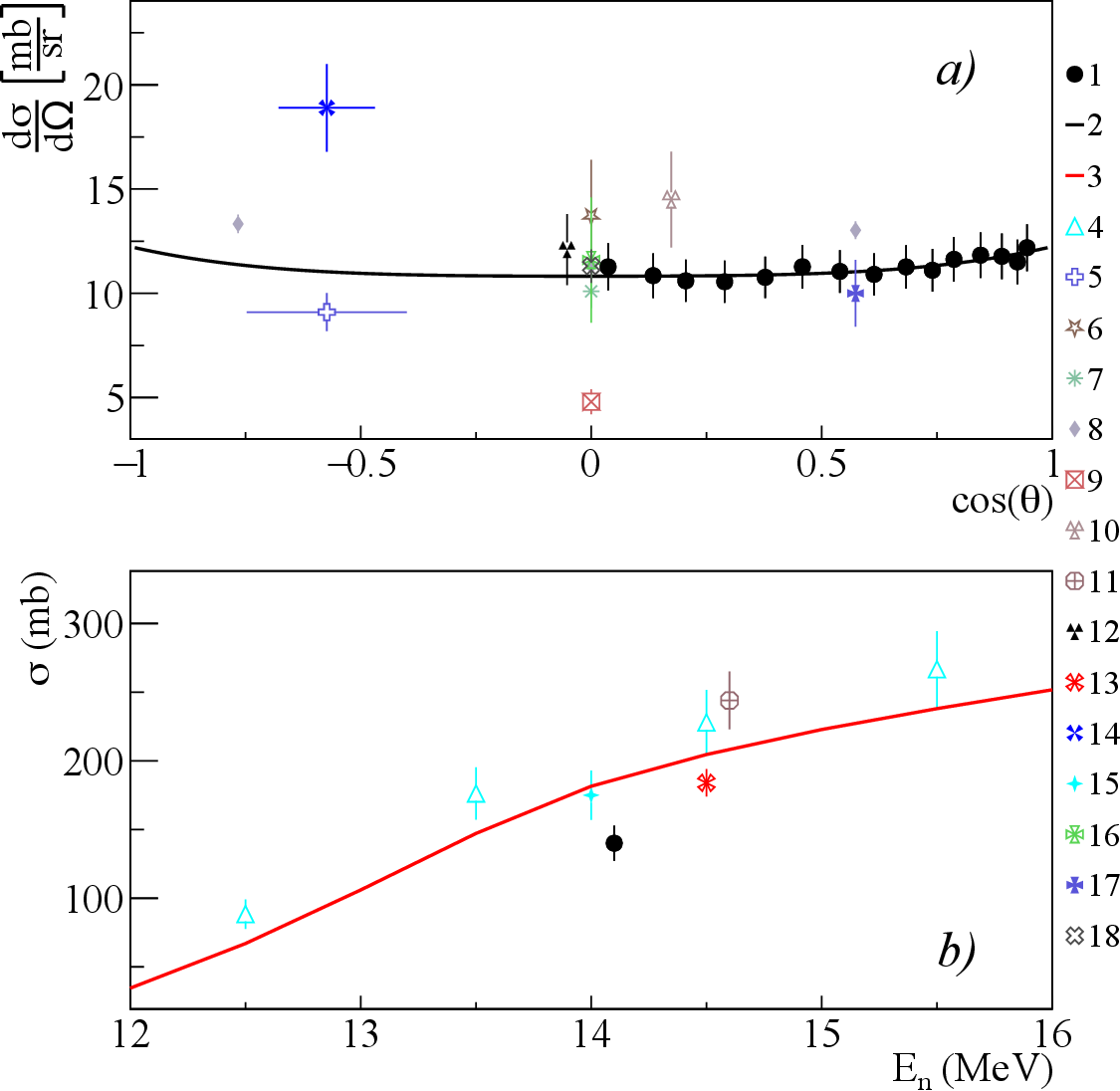}
\figcaption{Differential (a) and total (b) cross-sections of 1.808 MeV $\gamma$-ray emission from the $^{27}$Al$(n,d)^{26}$Mg reaction, compared with experimental data from other authors and theoretical calculations using the TALYS code with default parameters. 1 -- present work data; 2 -- Legendre polynomial approximation of angular distribution from present work; 3 -- TALYS calculation; 4 -- \cite{Pavlik1998}; 5 -- \cite{Hoot1975}; 6 -- \cite{Engesser1967}; 7 -- \cite{Clayeux1969}; 8 -- \cite{Zhou1997}; 9 -- \cite{Hasegawa1991}; 10 -- \cite{Nyberg1971}; 11 -- \cite{Hlavac1999}; 12 -- \cite{Zong1979}; 13 -- \cite{Simakov1998}; 14 -- \cite{Yamamoto1978}; 15 -- \cite{Bezotosnyi1975}; 16 -- \cite{Bochkarev1965}; 17 -- \cite{HINO1978}; 18 -- \cite{Hongyu1986}.}
\label{fig:fig6}
For the aluminum sample, angular distributions and total emission cross-sections were obtained for $\gamma$-ray lines with energies of $0.091$, $0.792$, $0.843$, $0.984$, $1.014$, $1.698$, $1.808$, $2.212$, $2.298$, $3.004$~MeV, generated in the reactions $^{27}$Al$(n,n')^{27}$Al, $^{27}$Al$(n,\alpha)^{24}$Na, $^{27}$Al$(n,p)^{27}$Mg, $^{27}$Al$(n,d)^{26}$Mg. It should be noted that the available literature data on differential cross-sections for the above lines \cite{Engesser1967,Clayeux1969,
Zong1979,Hasegawa1991,Prud'homme1960,Arya1967,Kinney1972,Yamamoto1978,Sukhanov1970} are extremely fragmentary and mostly limited to 1-2 angular data-points. Data on the total emission cross-sections of these lines are also limited to a small number of studies \cite{Bezotosnyi1975,Murata1988,Lashuk1994,
Burymov1969,Pavlik1998,Hlavac1999}, which show significant scatter, reaching in some cases $30-50$\,\% (see Figs.~\ref{fig:fig6} and \ref{fig:fig7}).

\vspace{12pt}
\includegraphics[width=75mm]{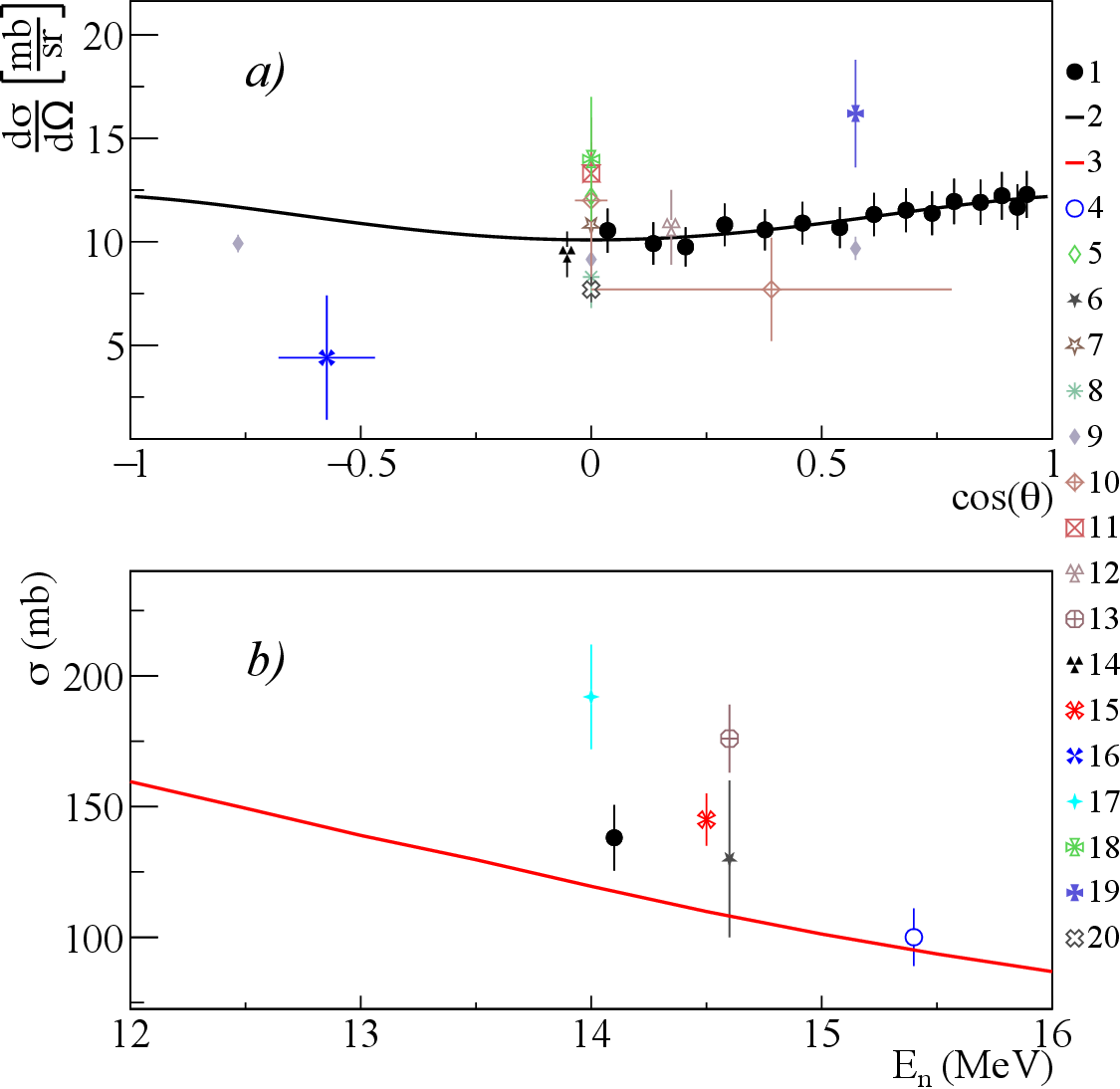}
\figcaption{Differential (a) and total (b) cross-sections of $2.212$~MeV $\gamma$-ray emission from the $^{27}$Al$(n,n')^{27}$Al reaction, compared with experimental data from other authors and theoretical calculations using the TALYS code with default parameters. 1 -- present work data; 2 -- Legendre polynomial approximation of angular distribution from present work; 3 -- TALYS calculation; 4 -- \cite{Lashuk1994}; 5 -- \cite{Sukhanov1970}; 6 -- \cite{Burymov1969}; 7 -- \cite{Engesser1967}; 8 -- \cite{Clayeux1969}; 9 -- \cite{Zhou1997}; 10 -- \cite{Prud'homme1960}; 11 -- \cite{Hasegawa1991}; 12 -- \cite{Nyberg1971}; 13 -- \cite{Hlavac1999}; 14 -- \cite{Zong1979}; 15 -- \cite{Simakov1998}; 16 -- \cite{Yamamoto1978}; 17 -- \cite{Bezotosnyi1975}; 18 -- \cite{Bochkarev1965}; 19 -- \cite{HINO1978}; 20 -- \cite{Hongyu1986}.}
\label{fig:fig7}

\end{multicols}

\newpage
\begin{center}
\tabcaption{\label{table:Al} Total emission cross-sections $\sigma$ and Legendre polynomial expansion coefficients $a_2$ and $a_4$ for angular distributions of $\gamma$-ray lines emitted during the interaction of $14.1$~MeV neutrons with aluminum nuclei. The energies of the initial ($i$) and final ($f$) states are given in MeV.}
\footnotesize
\begin{tabular}{ c c c c c c }
\hline
$E_{\gamma}$~(MeV) & Reaction     & Transition, $E_{i}$ $(J_{i}^{\pi}) \rightarrow E_{f} (J_{f}^{\pi})$ & $\sigma$~(mb) & $a_2$      & $a_4$         \\ \hline
 \hline
$0.091$ & $^{27}$Al$(n,\alpha)^{24}Na$ & $0.563(2^+)\rightarrow0.472(1^+)$      & $42\pm4$   & $-0.21\pm0.04$ & $0.03\pm0.06$   \\ \hline
$0.792$ & $^{27}$Al$(n,n')^{27}$Al                                                          & $3.004(9/2^+)\rightarrow 2.212(7/2^+)$  & $26\pm3$   & $-0.07\pm0.07$ & $-0.02\pm0.10$  \\ \hline
$0.843$ & $^{27}$Al$(n,n')^{27}$Al                                                          & $0.843(1/2^+)\rightarrow $g.s.$(5/2^+)$   & $28\pm3$   & $-0.24\pm0.07$ & $0.11\pm0.10$   \\ \hline
$0.984$ & $^{27}$Al$(n,p)^{27}$Mg                                                           & $0.984(3/2^+)\rightarrow $g.s.$(1/2^+)$   & $28\pm3$   & $0.07\pm0.09$  & $-0.07\pm0.14$  \\ \hline
$1.014$ & $^{27}$Al$(n,n')^{27}$Al                                                          & $1.014(3/2^+)\rightarrow $g.s.$(5/2^+)$   & $70\pm6$   & $0.04\pm0.04$  & $0.04\pm0.06$   \\ \hline
$1.698$ & $^{27}$Al$(n,p)^{27}$Mg                                                           & $1.698(5/2^+)\rightarrow $g.s.$(1/2^+)$   & $29\pm3$   & $0.23\pm0.15$  & $0.18\pm0.20$   \\ \hline
$1.808$ & $^{27}$Al$(n,d)^{26}$Mg                                                           & $1.808(2^+)\rightarrow $g.s.$(0^+)$       & $140\pm13$ & $0.08\pm0.01$  & $-0.02\pm0.02$  \\ \hline
$2.212$ & $^{27}$Al$(n,n')^{27}$Al                                                          & $2.212(7/2^+)\rightarrow $g.s.$(5/2^+)$   & $138\pm12$ & $0.14\pm0.02$  & $-0.03\pm0.02$  \\ \hline
$2.298$ & $^{27}$Al$(n,n')^{27}$Al                                                          & $4.510(11/2^+)\rightarrow 2.212(7/2^+)$ & $29\pm3$   & $0.34\pm0.04$  & $-0.03\pm0.06$  \\ \hline
$3.004$ & $^{27}$Al$(n,n')^{27}$Al                                                          & $3.004(9/2^+)\rightarrow $g.s.$(5/2^+)$   & $108\pm10$ & $0.20\pm0.02$  & $0.03\pm0.03$   \\ \hline
\end{tabular}
\end{center}

\begin{multicols}{2}

For silicon, data on angular distributions and total emission cross-sections were obtained for $\gamma$-ray lines with energies of $0.389$, $0.585$, $1.622$, $1.779$, $2.271$, $2.839$~MeV, generated in the reactions $^{28}$Si$(n,p)^{28}$Al, $^{28}$Si$(n,\alpha)^{25}$Mg, $^{28}$Si$(n,n')^{28}$Si, and $^{29}$Si$(n,2n)^{28}$Si.

\vspace{12pt}
\includegraphics[width=75mm]{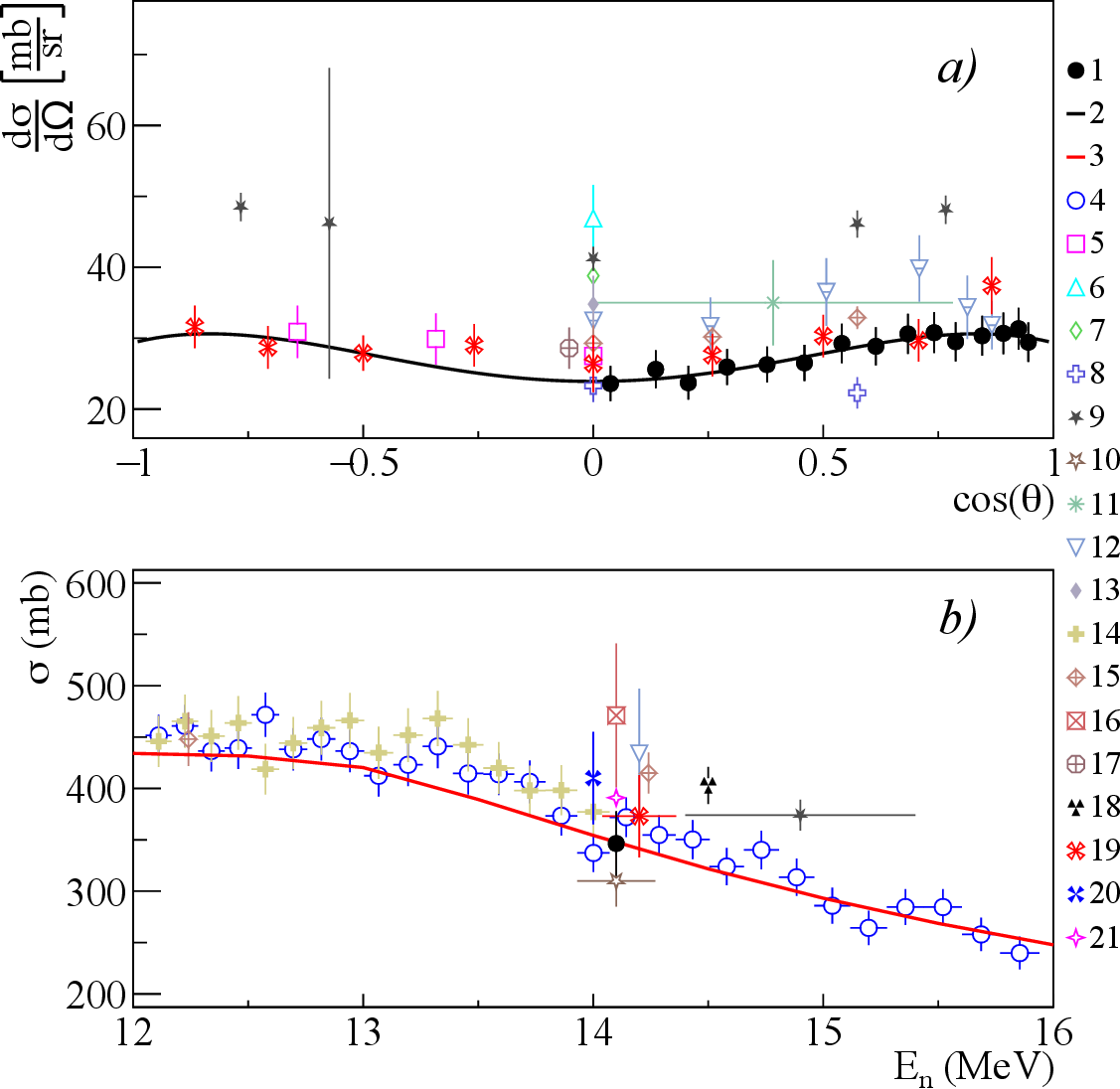}
\figcaption{Differential (a) and total (b) cross-sections of $1.779$~MeV $\gamma$-ray emission from the $^{28}$Si$(n,n')^{28}$Si and $^{29}$Si$(n,2n)^{28}$Si reactions, compared with experimental data from other authors and theoretical calculations using the TALYS code with default parameters. 1 -- present work data; 2 -- Legendre polynomial approximation of angular distribution from present work; 3 -- TALYS calculation; 4 -- \cite{Negret2013}; 5 -- \cite{Drake1978}; 6 -- \cite{Engesser1967}; 7 -- \cite{Guoying1992}; 8 -- \cite{Grenier1974}; 9 -- \cite{Zhou2011}; 10 -- \cite{Murata1988}; 11 -- \cite{Prud'homme1960}; 12 -- \cite{Connell1975}; 13 -- \cite{Hasegawa1991}; 14 -- \cite{Boromiza2020}; 15 -- \cite{Drosg2002}; 16 -- \cite{Martin1965}; 17 -- \cite{Zong1979} ; 18 -- \cite{Simakov1998}; 19 -- \cite{Abbondanno1973}; 20 -- \cite{Bezotosnyj1980}; 21 -- \cite{Kopatch2025}.}
\label{fig:fig8}

 Data for the most intense lines corresponding to the transitions $1.779(2^+) \rightarrow $g.s.$(5/2^+)$ and $4.617(4^+) \rightarrow 1.779(2^+)$ in the $^{28}$Si nucleus are presented in Figs.~\ref{fig:fig8} and \ref{fig:fig9} in comparison with data from the literature \cite{Murata1988,Hlavac1999,Boromiza2020,Martin1965,Abbondanno1973,Connell1975,Bezotosnyj1980,Drosg2002,Zhou2011,
Negret2013,Guoying1992}. As can be seen from the figures, the data from the present work for these lines agree with the data from other authors within the measurement uncertainties, both for angular distributions and total cross-sections. The results presented for silicon in the previous work within the TANGRA project \cite{Kopatch2025} differ somewhat from the results presented in this work. 

\vspace{12pt}
\includegraphics[width=75mm]{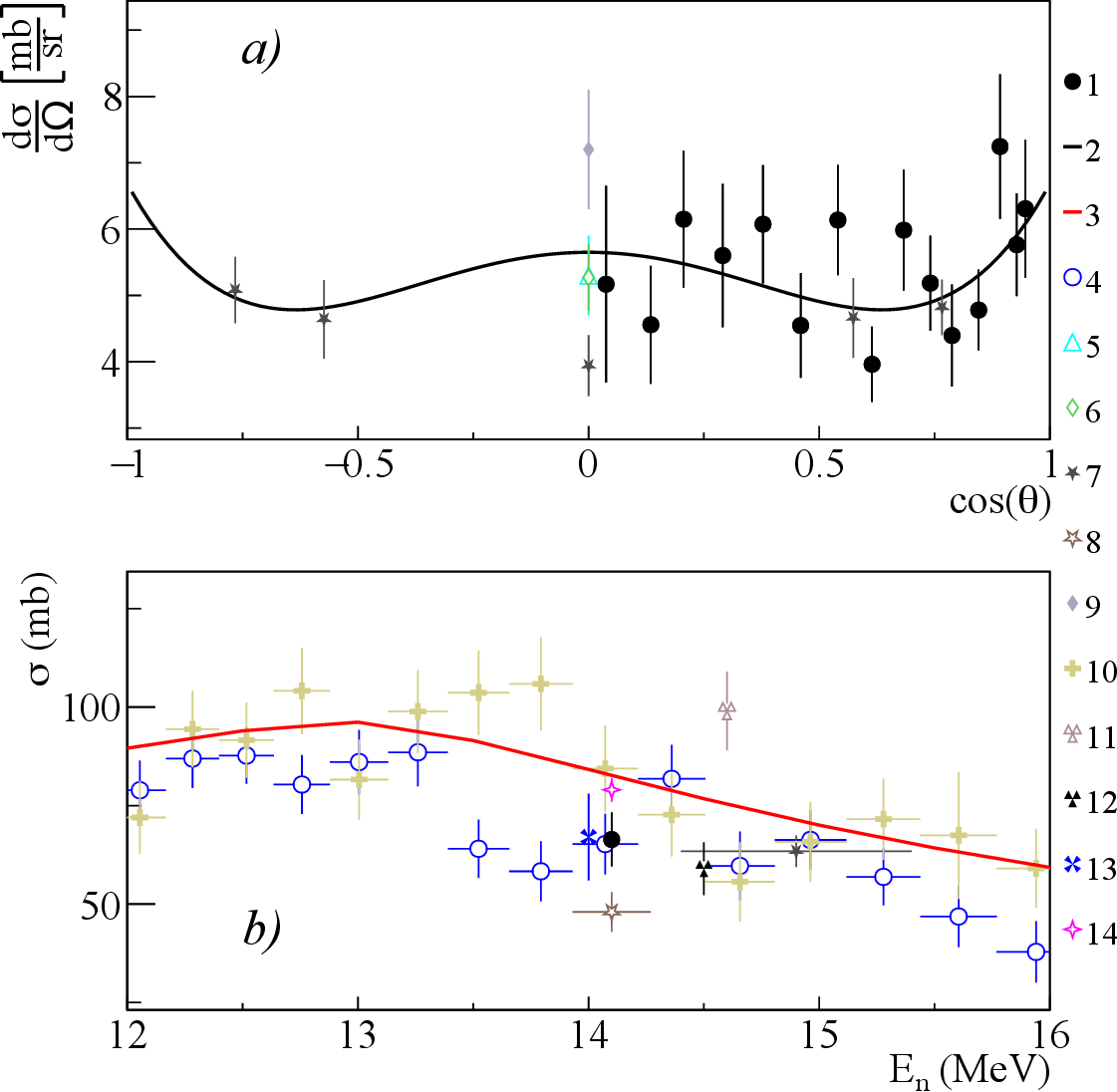}
\figcaption{Differential (a) and total (b) cross-sections of $2.839$~MeV $\gamma$-ray emission from the $^{28}$Si$(n,n')^{28}$Si reactions, compared with experimental data from other authors and theoretical calculations using the TALYS code with default parameters. 1 -- present work data; 2 -- Legendre polynomial approximation of angular distribution from present work; 3 -- TALYS calculation; 4 -- \cite{Negret2013}; 5 -- \cite{Engesser1967}; 6 -- \cite{Guoying1992}; 7 -- \cite{Zhou2011}; 8 -- \cite{Murata1988}; 9 -- \cite{Hasegawa1991}; 10 -- \cite{Boromiza2020}; 11 -- \cite{Hlavac1999}; 12 -- \cite{Simakov1998}; 13 -- \cite{Bezotosnyj1980}; 14 -- \cite{Kopatch2025}.}
\label{fig:fig9}

It should be noted that the data from \cite{Kopatch2025} were preliminary, as the cross-section calculation did not account for corrections for neutron multiple scattering and neutron attenuation in the generator wall. At the same time, only statistical uncertainty was considered in the error calculation in \cite{Kopatch2025}.

Analysis of the measured angular distributions showed a significant contribution from the Legendre polynomial expansion coefficients $a_2$ and $a_4$ for the $1.779$~MeV line (transition $1.779(2^+) \rightarrow $g.s.$(5/2^+$) in the $^{28}$Si nucleus). For the $1.622$ and $2.839$~MeV lines (transitions $1.622(2^+) \rightarrow $g.s.$(3^+)$ in the $^{28}$Al nucleus and $4.617(4^+) \rightarrow 1.779(2^+)$ in the $^{28}$Si nucleus), the contribution of the $a_2$ coefficient is negligible, while the $a_4$ coefficient is significant. For transitions in the $^{25}$Mg nucleus ($0.974(3/2^+) \rightarrow 0.585(1/2^+)$ and $0.585(1/2^+) \rightarrow $g.s.$(5/2^+)$), the errors in the coefficients were comparable to their magnitudes due to the large statistical scatter in the data.

\end{multicols}

\begin{center}
\tabcaption{\label{table:Si} Total emission cross-sections $\sigma$ and Legendre polynomial expansion coefficients $a_2$ and $a_4$ for angular distributions of $\gamma$-ray lines emitted during the interaction of $14.1$~MeV neutrons with silicon nuclei. The energies of the initial ($i$) and final ($f$) states are given in MeV.}
\footnotesize
\begin{tabular}{ c c c c c c }
\hline
$E_{\gamma}$~(MeV) & Reaction     & Transition, $E_{i}$ $(J_{i}^{\pi}) \rightarrow E_{f} (J_{f}^{\pi})$ & $\sigma$~(mb) & $a_2$      & $a_4$         \\ \hline
 \hline
$0.389$ & $^{28}$Si$(n,\alpha)^{25}$Mg                                                           & $0.974(3/2^+)\rightarrow 0.585(1/2^+)$  & $32\pm5$   & $-0.25\pm0.25$ & $-0.08\pm0.35$ \\ \hline
$0.585$ & $^{28}$Si$(n,\alpha)^{25}$Mg                                                           & $0.585(1/2^+)\rightarrow $g.s.$(5/2^+)$   & $35\pm4$   & $0.10\pm0.16$  & $0.16\pm0.21$   \\ \hline
$1.622$ & $^{28}$Si$(n,p)^{28}$Al                                                           & $1.622(2^+)\rightarrow $g.s.$(3^+)$       & $43\pm5$   & $0.01\pm0.18$  & $-0.27\pm0.25$  \\ \hline
$1.779$ & \begin{tabular}[c]{@{}l@{}} $^{28}$Si$(n,n')^{28}$Si \\ $^{29}$Si$(n,2n)^{28}$Si\end{tabular} & $1.779(2^+)\rightarrow $g.s.$(5/2^+)$     & $346\pm31$ & $0.18\pm0.02$  & $-0.11\pm0.03$  \\ \hline
$2.271$ & $^{28}$Si$(n,p)^{28}$Al                                                           & $2.271(4^+)\rightarrow $g.s.$(3^+)$       & $47\pm14$  & $0.57\pm0.70$  & $0.54\pm0.83$   \\ \hline
$2.839$ & $^{28}$Si$(n,n')^{28}$Si                                                          & $4.617(4^+)\rightarrow 1.779(2^+)$      & $67\pm7$   & $0.03\pm0.10$  & $0.23\pm0.14$ \\ \hline
\end{tabular}
\end{center}
\vspace{12pt}

\begin{multicols}{2}

\subsection{Calcium and Titanium}
\label{Calcium and Titanium}
The measurement results for the most intense $\gamma$-ray lines emitted during the interaction of $14.1$~MeV neutrons with calcium and titanium nuclei are presented in Figures~\ref{fig:fig10}\,--\,\ref{fig:fig13} and in Table~\ref{table:Ca} and \ref{table:Ti}. 

In the case of calcium, it should be noted that nearly all available experimental data are limited to measurements at $90^{\circ}$ (see, for example, \cite{Engesser1967}), and there are no reliable data whatsoever on total emission cross-sections or angular distributions. In the present work, detailed angular distributions were obtained for the first time for lines with energies of $0.770$, $0.891$, $2.814$, $3.736$, and $3.904$~MeV, and their total emission cross-sections were determined (see Table~\ref{table:Ca}). The results for the most intense lines $0.891$ and $3.736$~MeV, are shown in Figure~\ref{fig:fig10} and \ref{fig:fig11}.

Analysis of the measured angular distributions revealed a significant contribution from the first three even-order Legendre polynomial expansion coefficients  $a_2$, $a_4$, and $a_6$  for the $2.814$ and $3.736$~MeV lines, which correspond to the transitions $2.814(7/2^-) \rightarrow $g.s.$(3/2^+)$ in the $^{39}$K nucleus and $3.736(3^-) \rightarrow $g.s.$(0^+)$ in $^{40}$Ca, respectively. Significant $a_2$ and $a_4$ coefficients, with no contribution from $a_6$, were observed for the transitions $0.800(2^-) \rightarrow 0.03(2^-)$ and $0.892(5^-) \rightarrow $g.s.$(4^-)$ in the $^{40}$K nucleus (lines with energies of $0.770$~MeV and $0.891$~MeV, respectively), as well as for $3.904(2^+) \rightarrow $g.s.$(0^+)$ (the $3.904$~MeV line) in the $^{40}$Ca nucleus.

\vspace{12pt}
\includegraphics[width=75mm]{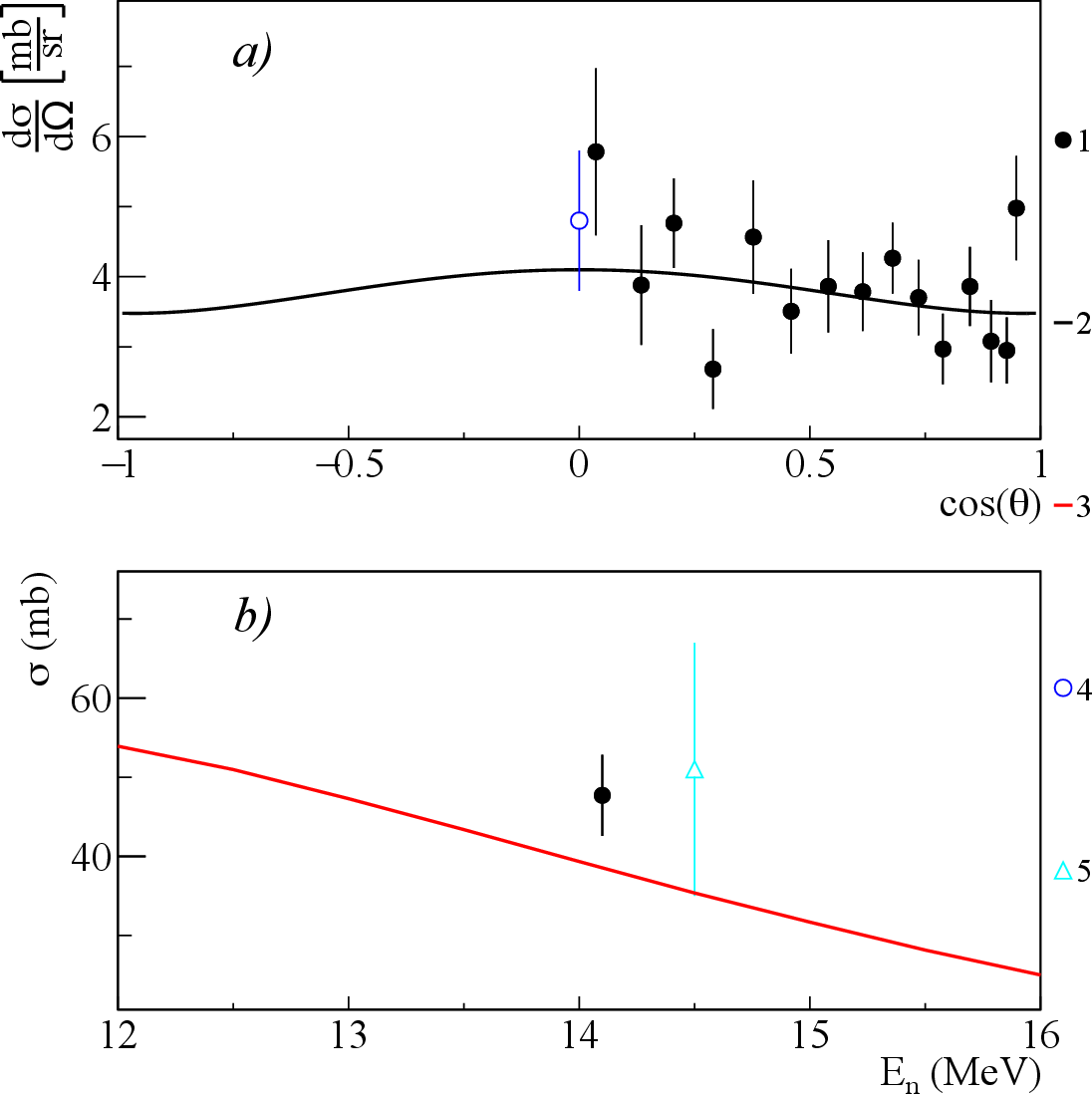}
\figcaption{Differential (a) and total (b) cross-sections of $0.891$~MeV $\gamma$-ray emission from the $^{40}$Ca$(n,p)^{40}$K reaction, compared with experimental data from other authors and theoretical calculations using the TALYS code with default parameters. 1 -- present work data; 2 -- Legendre polynomial approximation of angular distribution from present work; 3 -- TALYS calculation; 4 -- \cite{Engesser1967}; 5 -- \cite{Simakov1998}.}
\label{fig:fig10}

\includegraphics[width=75mm]{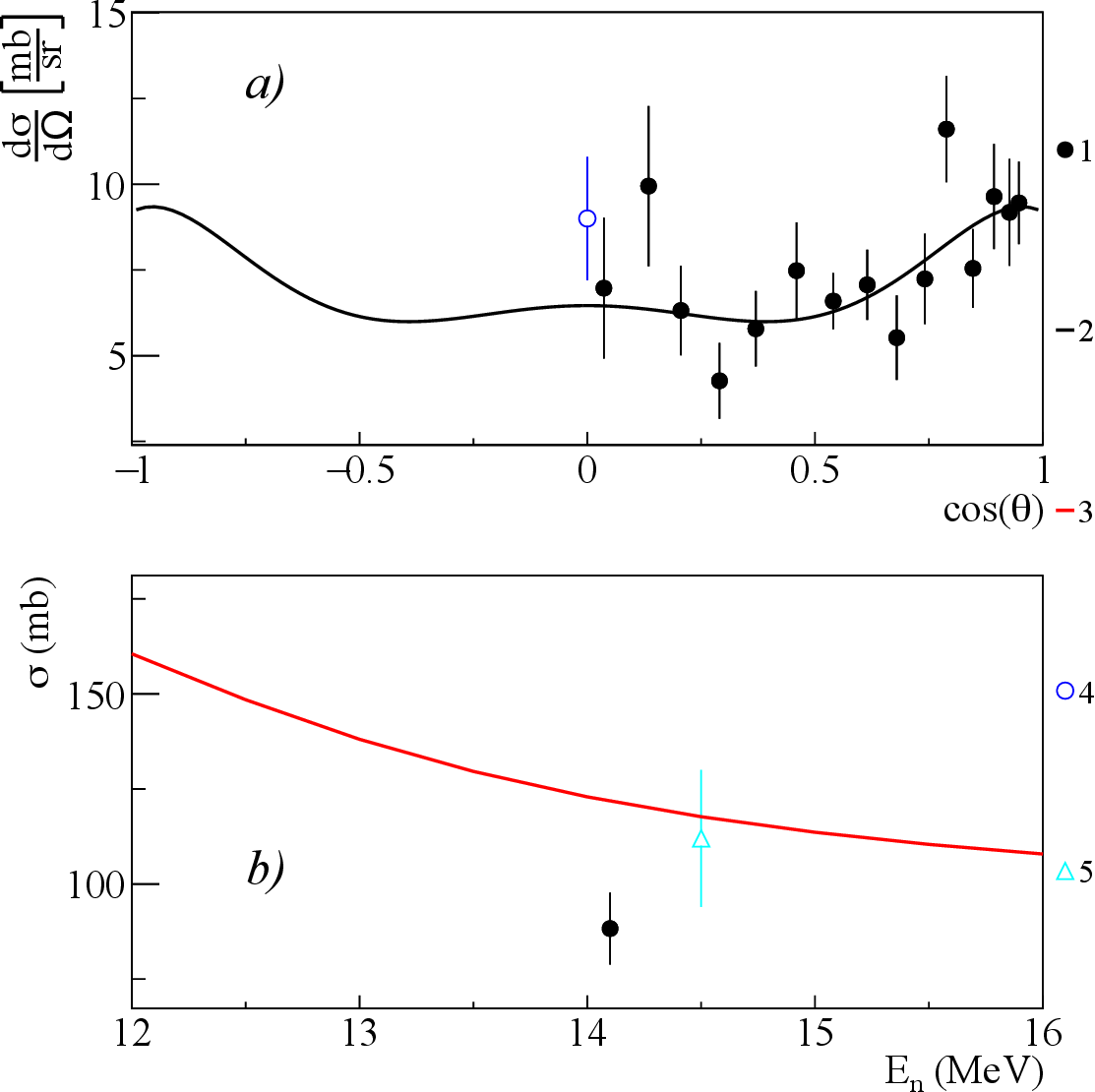}
\figcaption{Differential (a) and total (b) cross-sections of $3.736$~MeV $\gamma$-ray emission from the $^{40}$Ca$(n,n')^{40}$Ca reaction, compared with experimental data from other authors and theoretical calculations using the TALYS code with default parameters. 1 -- present work data; 2 -- Legendre polynomial approximation of angular distribution from present work; 3 -- TALYS calculation; 4 -- \cite{Engesser1967}; 5 -- \cite{Simakov1998}.}
\label{fig:fig11}

For titanium, in the present work, angular distributions and emission cross-sections were obtained for lines with energies of $0.121$, $0.130$, $0.159$, $0.175$, $0.370$, $0.889$, $0.944$, $0.983$, $1.037$, $1.120$, $1.312$, $1.437$, $1.542$, $2.375$~MeV. As can be seen from Figs.~\ref{fig:fig12} and \ref{fig:fig13}, which present data for the most intense lines with energies of $0.983$ and $1.312$~MeV in comparison with data from other authors, the total emission cross-sections obtained in the present work agree with most of the data from other authors \cite{Bezotosnyj1980,Breunlich1971,Dashdorj2005,Dashdorj2007} within the measurement uncertainties, except for data from a recent study \cite{Olacel2017}, which are slightly higher.

The available data on angular distributions for these and other lines are quite fragmentary and represented by only a few measurements \cite{Engesser1967,Abbondanno1973,Connell1975,Arya1967}, of which only two datasets provide more than 1 angular point \cite{Engesser1967,
Abbondanno1973}. Detailed angular distributions for all lines, except $0.983$ and $1.312$~MeV, were obtained for the first time.

\vspace{12pt}
\includegraphics[width=75mm]{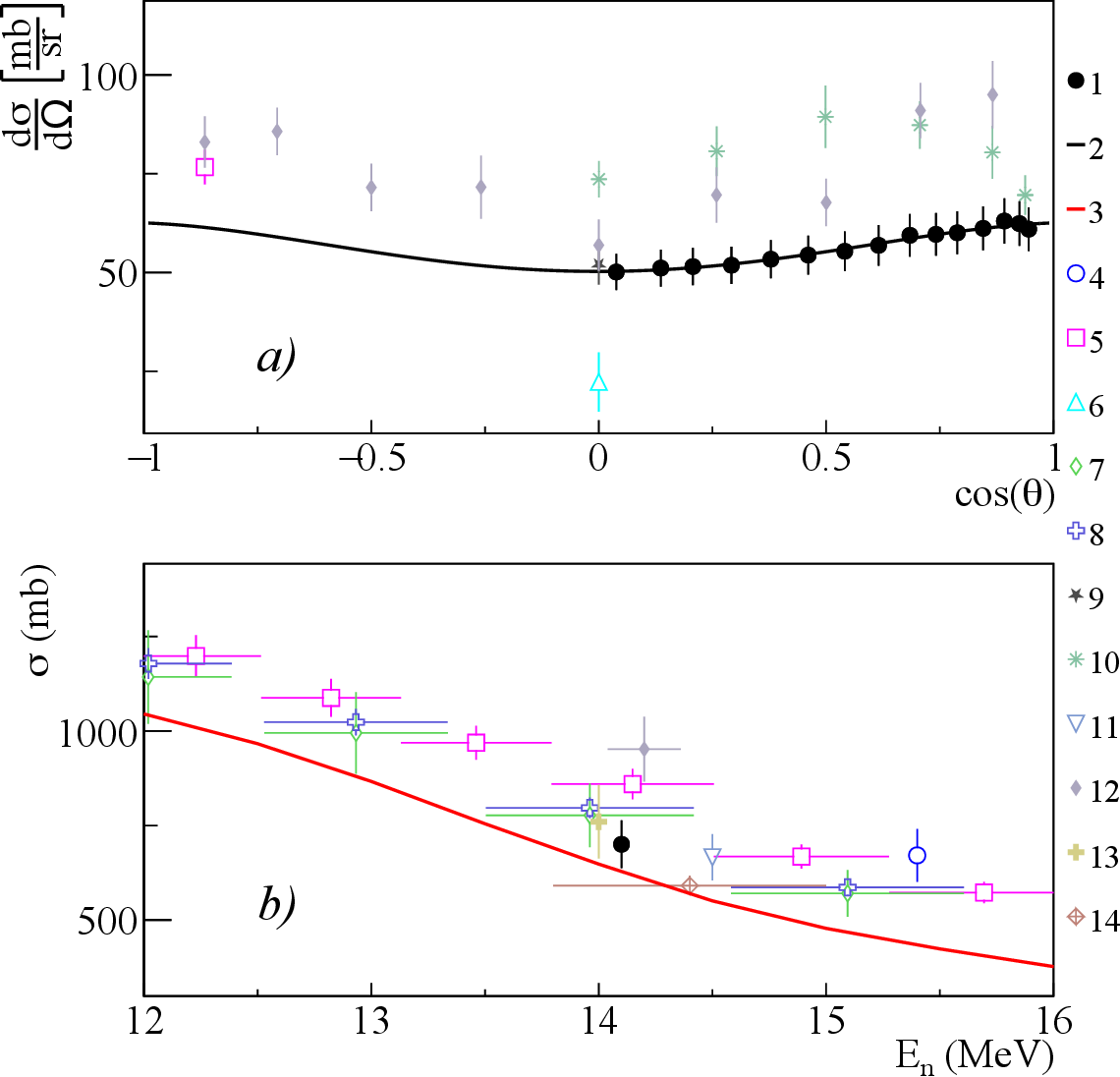}
\figcaption{Differential (a) and total (b) cross-sections of $0.983$~MeV $\gamma$-ray emission from the $^{48}$Ti$(n,n')^{48}$Ti and $^{49}$Ti$(n,2n)^{48}$Ti reactions, compared with experimental data from other authors and theoretical calculations using the TALYS code with default parameters. 1 -- present work data; 2 -- Legendre polynomial approximation of angular distribution from present work; 3 -- TALYS calculation; 4 -- \cite{Lashuk1994}; 5 -- \cite{Olacel2017}; 6 -- \cite{Arya1967}; 7 -- \cite{Dashdorj2005}; 8 -- \cite{Dashdorj2007}; 9 -- \cite{Engesser1967}; 10 -- \cite{Connell1975}; 11 -- \cite{Simakov1998}; 12 -- \cite{Abbondanno1973}; 13 -- \cite{Bezotosnyj1980}; 14 -- \cite{Breunlich1971}.}
\label{fig:fig12}
\end{multicols}

\begin{center}
\tabcaption{\label{table:Ca} Total emission cross sections $\sigma$ and coefficients of angular distribution expansion in Legendre polynomials $a_2, a_4$ and $a_6$ for $\gamma$-ray lines emitted in the interaction of $14.1$~MeV neutrons with calcium nuclei. The energies of the initial ($i$) and final ($f$) states are given in MeV.}
\footnotesize
\begin{tabular}{ c c c c c c c}
\hline
$E_{\gamma}$~(MeV) & Reaction     & Transition, $E_{i}$ $(J_{i}^{\pi}) \rightarrow E_{f} (J_{f}^{\pi})$ & $\sigma$~(mb) & $a_2$ & $a_4$ & $a_6$         \\ \hline
 \hline
$0.770$ & $^{40}$Ca$(n,p)^{40}$K                                                            & $0.800(2^-) \rightarrow 0.030(2^-)$     & $40\pm5$   & $0.10\pm0.21$  & $0.16\pm0.25$  &            \\ \hline
$0.891$ & $^{40}$Ca$(n,p)^{40}$K                                                            & $0.891(5^-) \rightarrow $g.s.$(4^-)$      & $48\pm5$   & $-0.13\pm0.12$ & $0.04\pm0.17$  &            \\ \hline
$1.159$ & $^{40}$Ca$(n,p)^{40}$K                                                            & $1.959(2^+) \rightarrow 0.800(2^-)$     & $29\pm4$   & $-0.07\pm0.19$ &            &            \\ \hline
$1.611$ & $^{40}$Ca$(n,\alpha)^{37}$Ar                                                           & $1.611(7/2^-) \rightarrow $g.s.$(3/2^+)$  & $34\pm6$   & $0.01\pm0.26$  & $-0.30\pm0.43$ & $0.41\pm0.47$  \\ \hline
$2.217$ & $^{40}$Ca$(n,\alpha)^{37}$Ar                                                           & $2.217(7/2^+) \rightarrow $g.s.$(3/2^+)$  & $21\pm3$   & $-0.38\pm0.20$ & $0.11\pm0.30$  &            \\ \hline
$2.814$ & $^{40}$Ca$(n,d)^{39}$K                                                            & $2.814(7/2^-) \rightarrow $g.s.$(3/2^+)$  & $28\pm4$   & $0.46\pm0.23$  & $-0.08\pm0.25$ & $0.49\pm0.34$  \\ \hline
$3.736$ & $^{40}$Ca$(n,n')^{40}$Ca                                                          & $3.736(3^-) \rightarrow $g.s$(0^+)$       & $88\pm9$   & $0.34\pm0.15$  & $0.12\pm0.18$  & $-0.15\pm0.24$ \\ \hline
$3.904$ & $^{40}$Ca$(n,n')^{40}$Ca                                                          & $3.904(2^+) \rightarrow $g.s.$(0^+)$      & $39\pm5$   & $0.07\pm0.22$  & $0.26\pm0.31$  &            \\ \hline
\end{tabular}
\end{center}
\begin{multicols}{2}

\includegraphics[width=75mm]{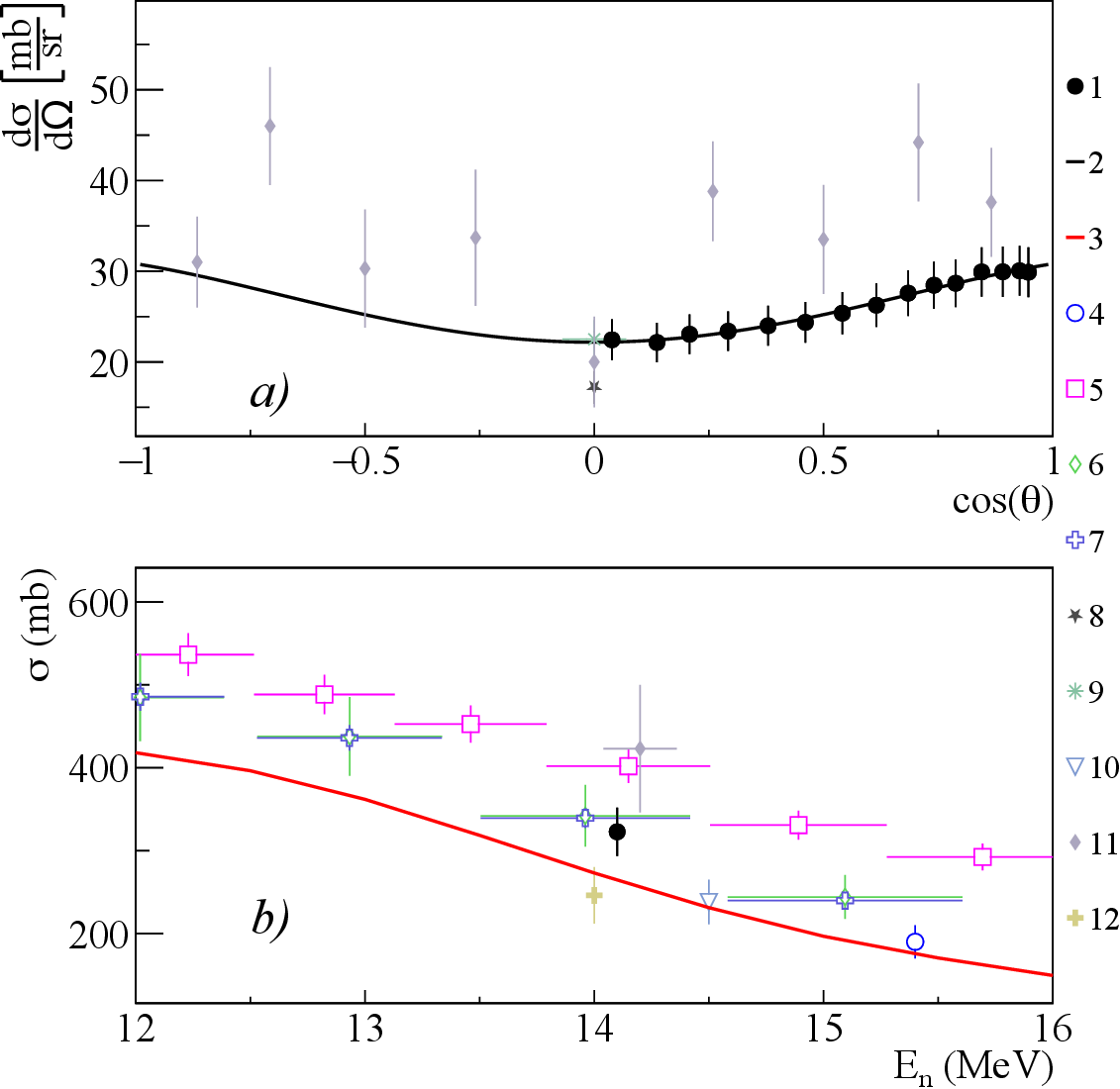}
\figcaption{Differential (a) and total (b) cross sections of $\gamma$-ray emission with energy $1.312$~MeV from the reactions $^{48}$Ti$(n,n')^{48}$Ti and $^{49}$Ti$(n,2n)^{48}$Ti, in comparison with experimental data of other authors and theoretical calculations based on the TALYS program with default parameters. 1 -- data from the present work; 2 -- angular distribution approximation from the present work using Legendre polynomials; 3 -- calculation in TALYS; 4 -- \cite{Lashuk1994}; 5 -- \cite{Olacel2017}; 6 -- \cite{Dashdorj2005}; 7 -- \cite{Dashdorj2007}; 8 -- \cite{Engesser1967}; 9 -- \cite{Connell1975}; 10 -- \cite{Simakov1998}; 11 -- \cite{Abbondanno1973}; 12 -- \cite{Bezotosnyj1980}.}
\label{fig:fig13}

Overall, as can be seen from the approximation results presented in Table~\ref{table:Ti}, for most transitions in titanium nuclei, significant $a_2$ coefficients were observed in the Legendre polynomial expansion, with negligible contributions from $a_4$ and $a_6$ coefficients. An exception is the transition $0.130(5^+) \rightarrow $g.s.$(6^+)$ in the $^{48}$Sc nucleus, for which the angular distribution is nearly isotropic.

\end{multicols}

\begin{center}
\tabcaption{\label{table:Ti} Total emission cross sections $\sigma$ and coefficients of angular distribution expansion in Legendre polynomials $a_2$ and $a_4$ for $\gamma$-ray lines emitted in the interaction of $14.1$~MeV neutrons with titanium nuclei. The energies of the initial ($i$) and final ($f$) states are given in MeV.}
\footnotesize
\begin{tabular}{ c c c c c c}
\hline
$E_{\gamma}$~(MeV) & Reaction     & Transition, $E_{i}$ $(J_{i}^{\pi}) \rightarrow E_{f} (J_{f}^{\pi})$ & $\sigma$~(mb) & $a_2$ & $a_4$        \\ \hline
 \hline
$0.121$ & $^{48}$Ti$(n,p)^{48}$Sc                                                           & $0.252(4^+) \rightarrow 0.130(5^+)$     & $51\pm8$   & $0.39\pm0.31$  & $-0.03\pm0.46$             \\ \hline
$0.130$ & $^{48}$Ti$(n,p)^{48}$Sc                                                           & $0.130(5^+) \rightarrow $g.s.$(6^+)$      & $47\pm5$   & $0.01\pm0.09$  & $-0.22\pm0.13$             \\ \hline
$0.159$ & \begin{tabular}[c]{@{}c@{}}$^{47}$Ti$(n,n')^{47}$Ti\\ $^{48}$Ti$(n,2n)^{47}$Ti\end{tabular} & $0.159(7/2^-) \rightarrow $g.s.$(5/2^-)$  & $187\pm17$ & $-0.27\pm0.01$ & $-0.03\pm0.02$             \\ \hline
$0.175$ & \begin{tabular}[c]{@{}c@{}}$^{48}$Ti$(n,n')^{48}$Ti\\ $^{49}$Ti$(n,2n)^{48}$Ti\end{tabular} & $3.508(6^+) \rightarrow 3.333(6^+)$     & $43\pm4$   & $0.42\pm0.05$  & $-0.12\pm0.07$             \\ \hline
$0.370$ & $^{48}$Ti$(n,p)^{48}$Sc                                                           & $0.622(3^+) \rightarrow 0.252(4^+)$     & $26\pm3$   & $0.15\pm0.08$  & $-0.07\pm0.11$             \\ \hline
$0.889$ & \begin{tabular}[c]{@{}c@{}}$^{46}$Ti$(n,n')^{46}$Ti\\ $^{47}$Ti$(n,2n)^{46}$Ti\end{tabular} & $0.889(2^+) \rightarrow $g.s.$(0^+)$      & $523\pm48$ & $0.12\pm0.03$  & $0.01\pm0.04$              \\ \hline
$0.944$ & \begin{tabular}[c]{@{}c@{}}$^{48}$Ti$(n,n')^{48}$Ti\\ $^{49}$Ti$(n,2n)^{48}$Ti\end{tabular} & $3.239(4^+) \rightarrow 2.295(4^+)$     & $65\pm6$   & $0.24\pm0.04$  & $0.12\pm0.06$              \\ \hline
$0.983$ & \begin{tabular}[c]{@{}c@{}}$^{48}$Ti$(n,n')^{48}$Ti\\ $^{49}$Ti$(n,2n)^{48}$Ti\end{tabular} & $0.983(2^+) \rightarrow $g.s.$(0^+)$      & $700\pm63$ & $0.16\pm0.01$  & $-0.04\pm0.01$             \\ \hline
$1.037$ & \begin{tabular}[c]{@{}c@{}}$^{48}$Ti$(n,n')^{48}$Ti\\ $^{49}$Ti$(n,2n)^{48}$Ti\end{tabular} & $3.333(6^+) \rightarrow 2.295(4^+)$     & $72\pm7$   & $0.42\pm0.04$  & $-0.01\pm0.06$             \\ \hline
$1.120$ & \begin{tabular}[c]{@{}c@{}}$^{46}$Ti$(n,n')^{46}$Ti\\ $^{47}$Ti$(n,2n)^{46}$Ti\end{tabular} & $2.009(4^+) \rightarrow 0.889(2^+)$     & $200\pm19$ & $0.29\pm0.05$  & $-0.04\pm0.06$             \\ \hline
$1.312$ & \begin{tabular}[c]{@{}c@{}}$^{48}$Ti$(n,n')^{48}$Ti\\ $^{49}$Ti$(n,2n)^{48}$Ti\end{tabular} & $2.295(4^+) \rightarrow 0.983(2^+)$     & $323\pm29$ & $0.24\pm0.01$  & $-0.04\pm0.01$             \\ \hline
$1.438$ & \begin{tabular}[c]{@{}c@{}}$^{48}$Ti$(n,n')^{48}$Ti\\ $^{49}$Ti$(n,2n)^{48}$Ti\end{tabular} & $2.421(2^+) \rightarrow 0.983(2^+)$     & $59\pm6$   & $0.15\pm0.06$  & $-0.01\pm0.08$            \\ \hline
$1.542$ & \begin{tabular}[c]{@{}c@{}}$^{49}$Ti$(n,n')^{49}$Ti\\ $^{49}$Ti$(n,2n)^{50}$Ti\end{tabular} & $1.542(11/2^-) \rightarrow $g.s.$(7/2^-)$ & $659\pm61$ & $0.21\pm0.05$  & $0.08\pm0.06$             \\ \hline
$2.240$ & \begin{tabular}[c]{@{}c@{}}$^{48}$Ti$(n,n')^{48}$Ti\\ $^{49}$Ti$(n,2n)^{48}$Ti\end{tabular} & $3.224(3^+) \rightarrow 0.983(2^+)$     & $36\pm3$   & $0.15\pm0.05$  & $-0.22\pm0.09$            \\ \hline
$2.375$ & \begin{tabular}[c]{@{}c@{}}$^{48}$Ti$(n,n')^{48}$Ti\\ $^{49}$Ti$(n,2n)^{48}$Ti\end{tabular} & $3.358(3^-) \rightarrow 0.983(2^+)$     & $89\pm8$   & $-0.12\pm0.02$ &                       \\ \hline
$2.633$ & \begin{tabular}[c]{@{}c@{}}$^{48}$Ti$(n,n')^{48}$Ti\\ $^{49}$Ti$(n,2n)^{48}$Ti\end{tabular} & $3.617(3^+) \rightarrow 0.983(2^+)$     & $21\pm2$   & $-0.19\pm0.13$ & $-0.01\pm0.18$             \\ \hline
\end{tabular}
\end{center}
\begin{multicols}{2}

\subsection{Chromium and Iron}
\label{Chromium and Iron}

The results of measurements for the most intense $\gamma$-ray lines emitted during the interaction of 14.1 MeV neutrons with chromium and iron nuclei are presented in Figs.~\ref{fig:fig14}--\ref{fig:fig17} and Table~\ref{table:Cr} and \ref{table:Fe}.

\vspace{12pt}

\includegraphics[width=75mm]{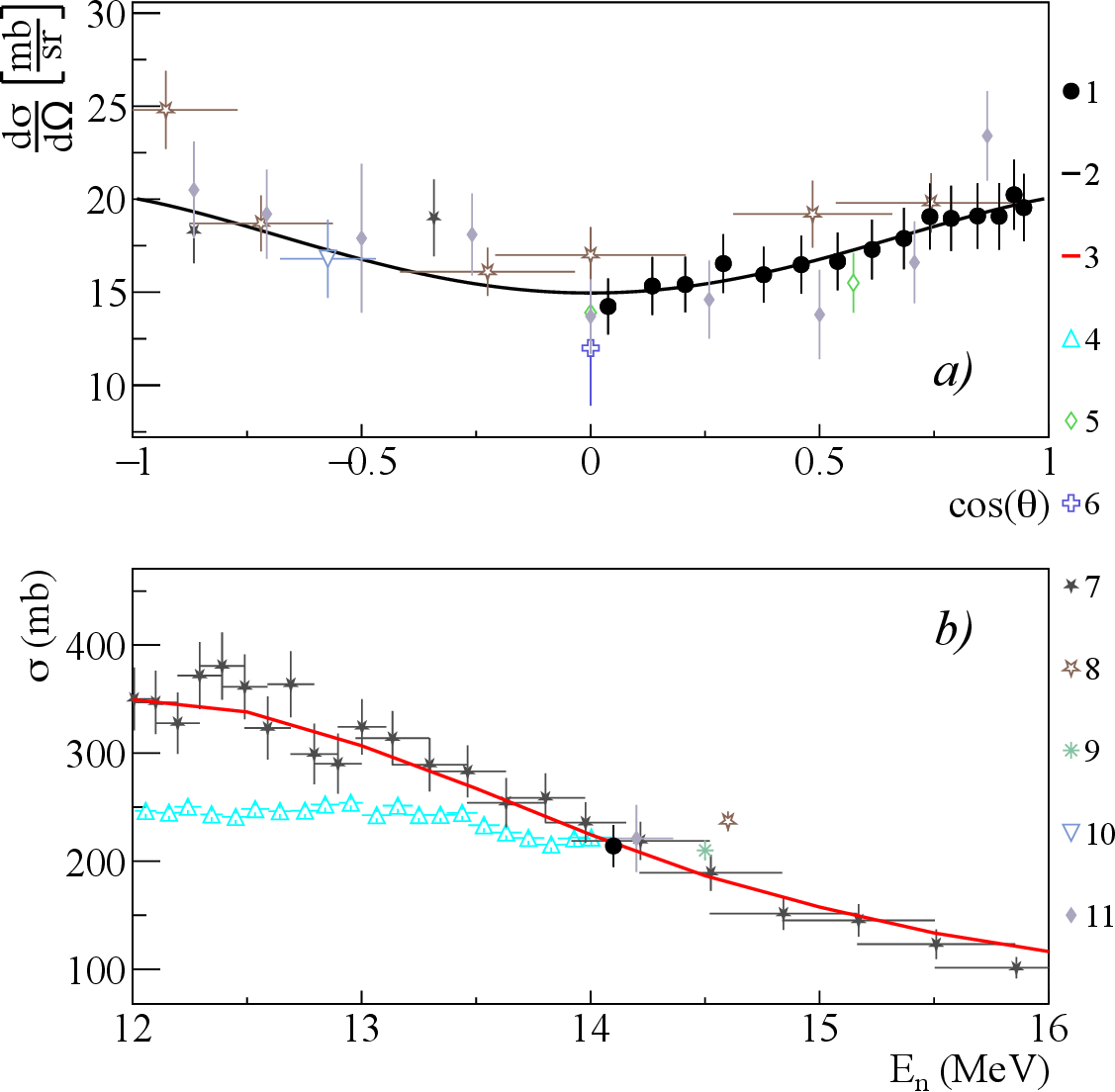}
\figcaption{Differential (a) and total (b) cross sections of $\gamma$-ray emission with energy $0.935$~MeV from the reactions $^{52}$Cr$(n,n')^{52}$Cr and $^{53}$Cr$(n,2n)^{52}$Cr, in comparison with experimental data of other authors and theoretical calculations based on the TALYS program with default parameters. 1 -- data from the present work; 2 -- angular distribution approximation from the present work using Legendre polynomials; 3 -- calculation in TALYS; 4 -- \cite{Voss1975}; 5 -- \cite{Clayeux1969}; 6 -- \cite{Kinney1972}; 7 -- \cite{Mihailescu2007}; 8 -- \cite{Oblozinsky1992}; 9 -- \cite{Simakov1998}; 10 -- \cite{Yamamoto1978}; 11 -- \cite{Abbondanno1973}.}
\label{fig:fig14}

\vspace{12pt}
For chromium, a relatively large amount of experimental data is available for both angular distributions and total emission cross sections. In this work, emission cross sections and angular distributions were obtained for $\gamma$-ray lines with energies of $0.647$, $0.704$, $0.744$, $0.935$, $1.333$, $1.434$, and $1.530$~MeV emitted in the reactions $^{52}$Cr$(n,n')^{52}$Cr and $^{53}$Cr$(n,2n)^{52}$Cr.

\vspace{12pt}
\includegraphics[width=75mm]{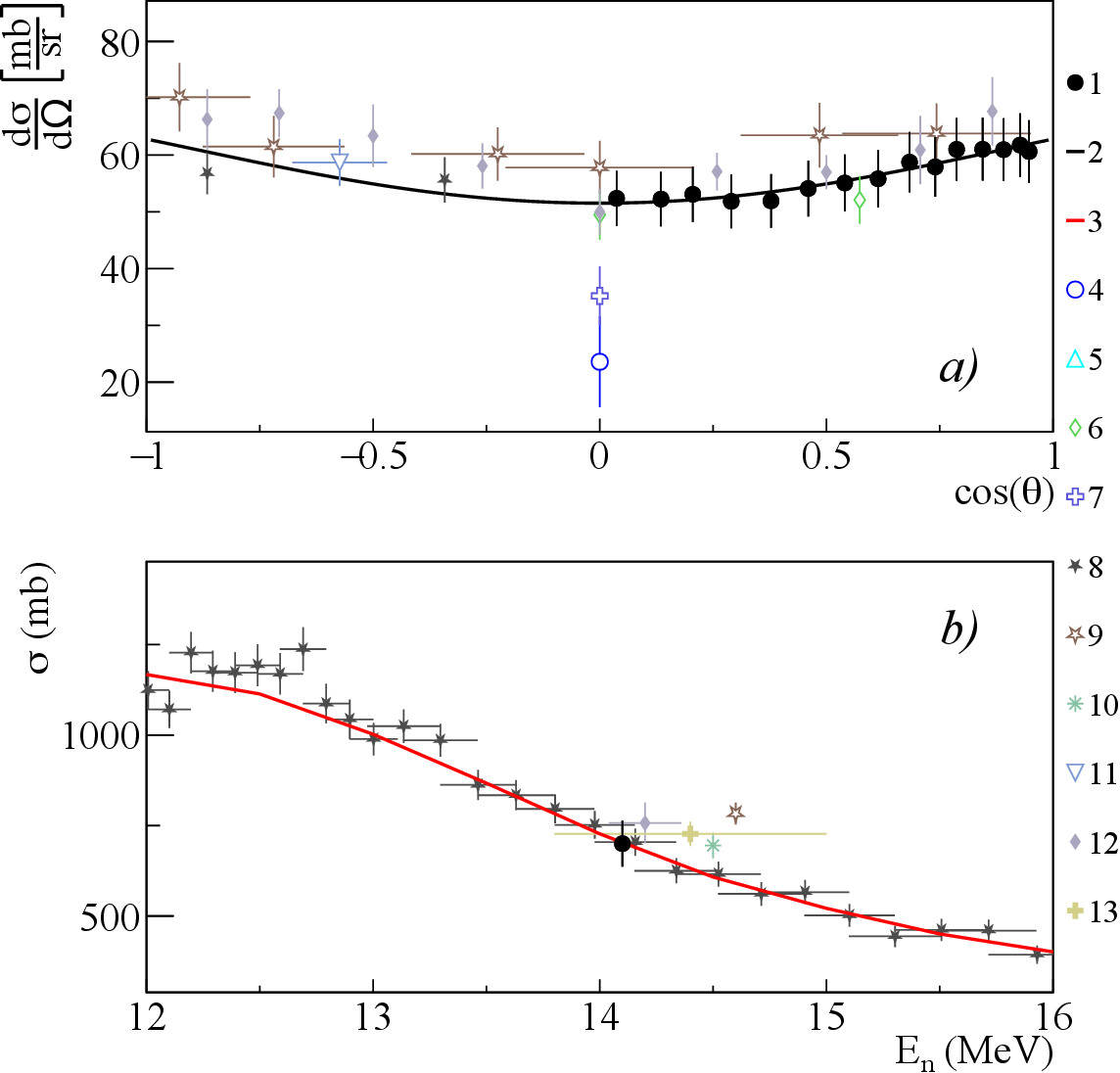}
\figcaption{Differential (a) and total (b) cross sections of $\gamma$-ray emission with energy $1.434$~MeV from the reactions $^{52}$Cr$(n,n')^{52}$Cr and $^{53}$Cr$(n,2n)^{52}$Cr, in comparison with experimental data of other authors and theoretical calculations based on the TALYS program with default parameters. 1 -- data from the present work; 2 -- angular distribution approximation from the present work using Legendre polynomials; 3 -- calculation in TALYS; 4 -- \cite{Arya1967}; 5 -- \cite{Voss1975}; 6 -- \cite{Clayeux1969}; 7 -- \cite{Kinney1972}; 8 -- \cite{Mihailescu2007}; 9 -- \cite{Oblozinsky1992}; 10 -- \cite{Simakov1998}; 11 -- \cite{Yamamoto1978}; 12 -- \cite{Abbondanno1973}; 13 -- \cite{Breunlich1971}.}
\label{fig:fig15}

\end{multicols}

\begin{center}
\tabcaption{\label{table:Cr} Total emission cross sections $\sigma$ and coefficients of angular distribution expansion in Legendre polynomials $a_2$ and $a_4$ for $\gamma$-ray lines emitted in the interaction of $14.1$~MeV neutrons with chromium nuclei. The energies of the initial ($i$) and final ($f$) states are given in MeV.}
\footnotesize
\begin{tabular}{ c c c c c c}
\hline
$E_{\gamma}$~(MeV) & Reaction     & Transition, $E_{i}$ $(J_{i}^{\pi}) \rightarrow E_{f} (J_{f}^{\pi})$ & $\sigma$~(mb) & $a_2$ & $a_4$ 
  \\ \hline
 \hline
$0.647$ & \begin{tabular}[c]{@{}c@{}}$^{52}$Cr$(n,n')^{52}$Cr\\ $^{53}$Cr$(n,2n)^{52}$Cr\end{tabular} & $3.415(4^+) \rightarrow 2.767(4^+)$                                                                  & $52\pm5$   & $0.25\pm0.06$  & $-0.02\pm0.08$ \\ \hline
$0.744$ & \begin{tabular}[c]{@{}c@{}}$^{52}$Cr$(n,n')^{52}$Cr\\ $^{53}$Cr$(n,2n)^{52}$Cr\end{tabular} & $3.113(6^+) \rightarrow 2.369(4^+)$                                                                  & $52\pm5$   & $0.39\pm0.11$  & $-0.21\pm0.15$ \\ \hline
$0.935$ & \begin{tabular}[c]{@{}c@{}}$^{52}$Cr$(n,n')^{52}$Cr\\ $^{53}$Cr$(n,2n)^{52}$Cr\end{tabular} & $2.369(4^+) \rightarrow 1.434(2^+)$                                                                  & $214\pm19$ & $0.22\pm0.02$  & $-0.04\pm0.02$ \\ \hline
$1.333$ & \begin{tabular}[c]{@{}c@{}}$^{52}$Cr$(n,n')^{52}$Cr\\ $^{53}$Cr$(n,2n)^{52}$Cr\end{tabular} & $2.767(4^+) \rightarrow 1.434(2^+)$                                                                  & $162\pm15$ & $0.19\pm0.01$  & $-0.05\pm0.02$ \\ \hline
$1.434$ & \begin{tabular}[c]{@{}c@{}}$^{52}$Cr$(n,n')^{52}$Cr\\ $^{53}$Cr$(n,2n)^{52}$Cr\end{tabular} & $1.434(2^+) \rightarrow $g.s.$(0^+)$                                                                   & $700\pm63$ & $0.14\pm0.01$  & $-0.01\pm0.02$ \\ \hline
$1.530$ & \begin{tabular}[c]{@{}c@{}}$^{52}$Cr$(n,n')^{52}$Cr\\ $^{53}$Cr$(n,2n)^{52}$Cr\end{tabular} & $2.964(2^+) \rightarrow 1.434(2^+)$                                                                  & $40\pm4$   & $-0.08\pm0.07$ & $-0.03\pm0.10$ \\ \hline
\end{tabular}
\end{center}
\begin{multicols}{2}

 For the most intense lines with energies of $0.935$ and $1.434$~MeV (see Figs.~\ref{fig:fig14} and \ref{fig:fig15}), agreement is observed between the results of this work and data from other authors \cite{Abbondanno1973,Arya1967,Breunlich1971,
Kinney1972,Grenier1974,Voss1975,Yamamoto1978,Oblozinsky1992,
Mihailescu2007}, both in angular distributions and total emission cross sections.

It should be noted that all measured angular distributions exhibit a pronounced $a_2$ coefficient in the Legendre polynomial expansion (see Table~\ref{table:Cr}), while the influence of the $a_4$ coefficient is negligible.

Data obtained for iron in the present work include angular distributions and total emission cross sections for $\gamma$-ray lines with energies of $0.125$, $0.212$, $0.846$, $0.931$, $1.037$, $1.238$, $1.303$, $1.408$, $1.670$, $1.810$, and $2.598$~MeV, generated in $(n,p)$, $(n,d)$, $(n,n')$, and $(n,2n)$ reactions on iron isotopes.

\vspace{12pt}
\includegraphics[width=75mm]{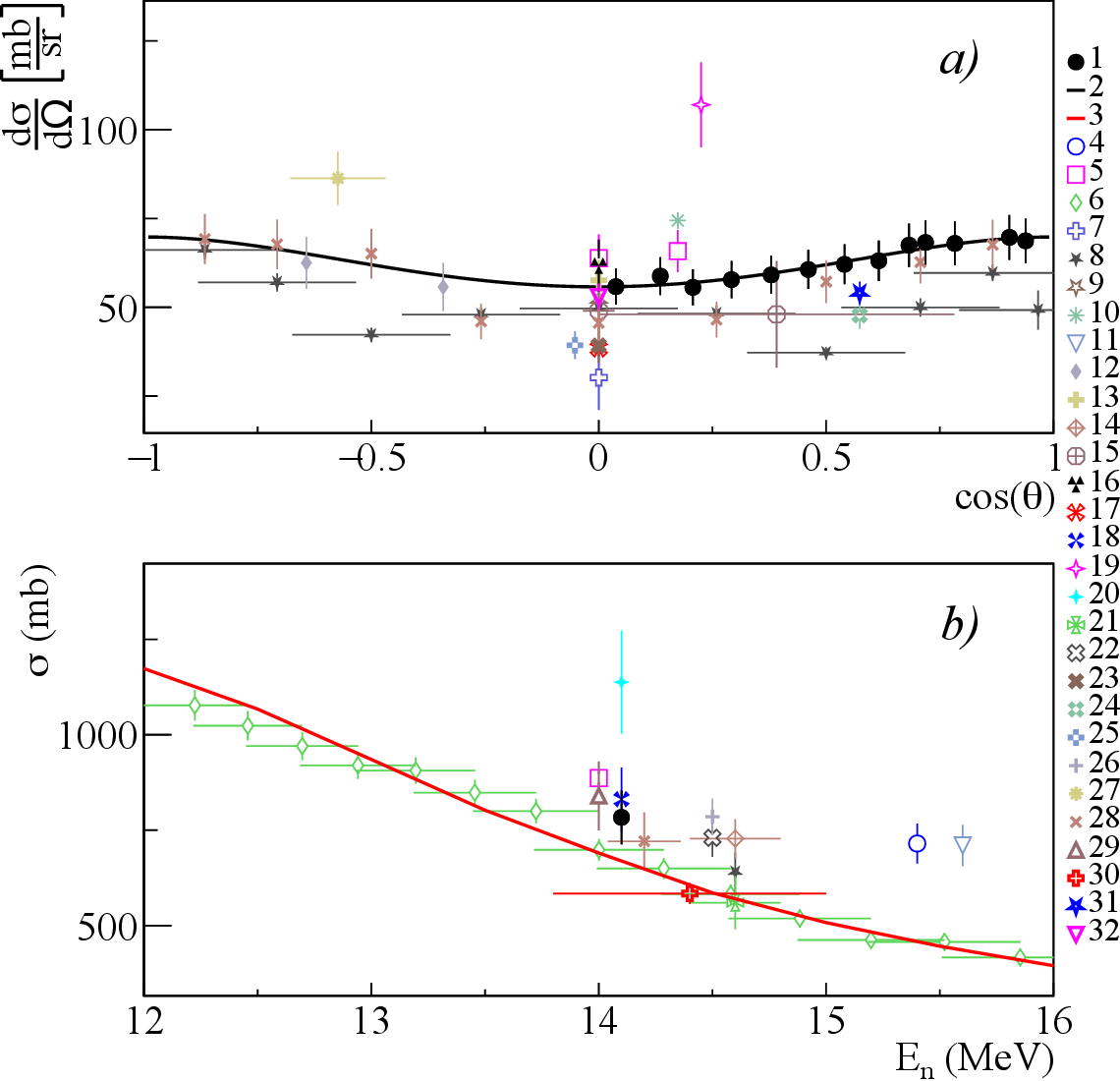}
\figcaption{Differential (a) and total (b) cross sections of $\gamma$-ray emission with energy $0.846$~MeV from the reactions $^{56}$Fe$(n,n')^{56}$Fe and $^{57}$Fe$(n,2n)^{56}$Fe, in comparison with experimental data of other authors and theoretical calculations based on the TALYS program with default parameters. 1 -- data from the present work; 2 -- angular distribution approximation from the present work using Legendre polynomials; 3 -- calculation in TALYS; 4 -- \cite{Lashuk1994}; 5 -- \cite{Shalabi1983}; 6 -- \cite{Negret2014}; 7 -- \cite{Arya1967}; 8 -- \cite{Degtyarev1977}; 9 -- \cite{Sukhanov1970}; 10 -- \cite{Joensson1969}; 11 -- \cite{Broder1970}; 12 -- \cite{Drake1978}; 13 -- \cite{Engesser1967}; 14 -- \cite{Western1965}; 15 -- \cite{Prud'homme1960}; 16 -- \cite{Hasegawa1991}; 17 -- \cite{Drosg2002}; 18 -- \cite{Mitsuda2002}; 19 -- \cite{Bostrom1959}; 20 -- \cite{Martin1965}; 21 -- \cite{Antalik1980}; 22 -- \cite{Nelson2005}; 23 -- \cite{Hlavac1983}; 24 -- \cite{Xiamin1982}; 25 -- \cite{Zong1979}; 26 -- \cite{Simakov1998}; 27 -- \cite{Yamamoto1978}; 28 -- \cite{Abbondanno1973}; 29 -- \cite{Bezotosnyi1975}; 30 -- \cite{Breunlich1971}; 31 -- \cite{Jinqiang1988}; 32 -- \cite{Hongyu1986}.}
\label{fig:fig16}

 Comparison of experimental data obtained in the present work with data from other authors \cite{Engesser1967,Martin1965,Bezotosnyi1975,Zong1979,Hasegawa1991,Lashuk1994,
Abbondanno1973,Drosg2002,Arya1967,Breunlich1971,Kinney1972,Yamamoto1978,Bostrom1959,Prud'homme1960,
Sukhanov1970,Degtyarev1977,Drake1978,Xiamin1982,Shalabi1983,Hlavac1983,Jinqiang1988,Negret2014,
Broder1970,Western1965,Mitsuda2002,Antalik1980,Nelson2005} for the most intense lines ($0.846$ and $1.238$~MeV) is shown in Figs.~\ref{fig:fig16} and \ref{fig:fig17}. As can be seen from the figures, there is a considerable amount of experimental data available for these lines, both for angular distributions and total cross sections. The values of both total and differential cross sections obtained in this work for these lines agree with data from other authors within the existing scatter of data points.

From the Legendre polynomial expansion coefficients presented in Table~\ref{table:Fe}, it is evident that for all transitions -- except for the $2.657(2^+) \rightarrow 0.846(2^+)$ transition in $^{56}$Fe ($1.810$~MeV line) and the $0.125(7/2^-) \rightarrow $g.s.$(5/2^-)$ transition in $^{55}$Mn ($0.125$~MeV line) -- there is a significant contribution from the $a_2$ coefficient, with a minor or negligible contribution from the $a_4$ coefficient.

\vspace{12pt}
\includegraphics[width=75mm]{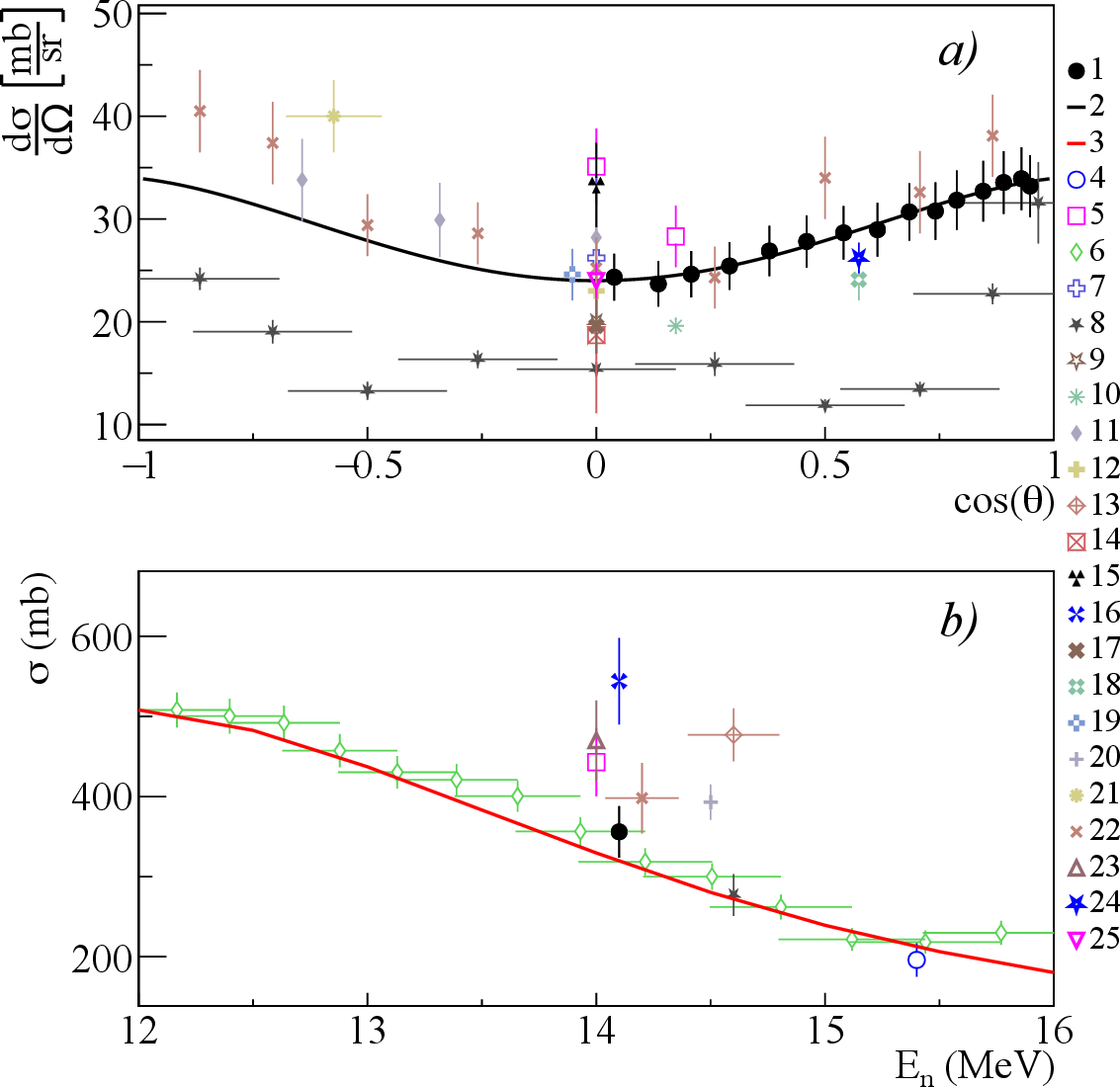}
\figcaption{Differential (a) and total (b) cross sections of $\gamma$-ray emission with energy $1.238$~MeV from the reactions $^{56}$Fe$(n,n')^{56}$Fe and $^{57}$Fe$(n,2n)^{56}$Fe, in comparison with experimental data of other authors and theoretical calculations based on the TALYS program with default parameters. 1 -- data from the present work; 2 -- angular distribution approximation from the present work using Legendre polynomials; 3 -- calculation in TALYS; 4 -- \cite{Lashuk1994}; 5 -- \cite{Shalabi1983}; 6 -- \cite{Negret2014}; 7 -- \cite{Arya1967}; 8 -- \cite{Degtyarev1977}; 9 -- \cite{Sukhanov1970}; 10 -- \cite{Joensson1969}; 11 -- \cite{Drake1978}; 12 -- \cite{Engesser1967}; 13 -- \cite{Western1965}; 14 -- \cite{Kinney1972}; 15 -- \cite{Hasegawa1991}; 16 -- \cite{Mitsuda2002}; 17 -- \cite{Hlavac1983}; 18 -- \cite{Xiamin1982}; 19 -- \cite{Zong1979}; 20 -- \cite{Simakov1998}; 21 -- \cite{Yamamoto1978}; 22 -- \cite{Abbondanno1973}; 23 -- \cite{Bezotosnyi1975}; 24 -- \cite{Jinqiang1988}; 25 -- \cite{Hongyu1986}.}
\label{fig:fig17}

\end{multicols}
\newpage
\begin{center}
\tabcaption{\label{table:Fe} Total emission cross sections $\sigma$ and coefficients of angular distribution expansion in Legendre polynomials $a_2$ and $a_4$ for $\gamma$-ray lines emitted in the interaction of $14.1$~MeV neutrons with iron nuclei. The energies of the initial ($i$) and final ($f$) states are given in MeV.}
\footnotesize
\begin{tabular}{ c c c c c c}
\hline
$E_{\gamma}$~(MeV) & Reaction     & Transition, $E_{i}$ $(J_{i}^{\pi}) \rightarrow E_{f} (J_{f}^{\pi})$ & $\sigma$~(mb) & $a_2$ & $a_4$ 
  \\ \hline
 \hline
$0.125$ & $^{56}$Fe$(n,d)^{55}$Mn                                                           & $0.125(7/2^-) \rightarrow $g.s.$(5/2^-)$                                                               & $46\pm4$   & $-0.11\pm0.04$ & $-0.19\pm0.06$ \\ \hline
$0.212$ & \begin{tabular}[c]{@{}c@{}}$^{56}$Fe$(n,p)^{56}$Mn\\ $^{57}$Fe$(n,d)^{56}$Mn\end{tabular}   & $0.212(4^+) \rightarrow $g.s.$(3^+)$                                                                   & $39\pm4$   & $-0.18\pm0.09$ & $0.07\pm0.12$  \\ \hline
$0.367$ & \begin{tabular}[c]{@{}c@{}}$^{56}$Fe$(n,n')^{56}$Fe\\ $^{57}$Fe$(n,2n)^{56}$Fe\end{tabular} & $3.756(6^+) \rightarrow 3.389(6^+)$                                                                  & $10\pm3$    & $0.16\pm0.60$  & $0.15\pm0.74$  \\ \hline
$0.411$ & \begin{tabular}[c]{@{}c@{}}$^{54}$Fe$(n,n')^{54}$Fe\\ $^{56}$Fe$(n,2n)^{55}$Fe\end{tabular} & \begin{tabular}[c]{@{}c@{}}$2.949(6^+) \rightarrow 2.538(4^+)$\\ $0.411(1/2^-) \rightarrow $g.s.$(3/2^-)$\end{tabular} & $20\pm2$   & $0.07\pm0.04$  & $0.02\pm0.05$  \\ \hline
$0.477$ & $^{56}$Fe$(n,2n)^{55}$Fe                                                          & $1.408(7/2^-) \rightarrow 0.931(5/2^-)$                                                              & $17\pm3$   & $-0.28\pm0.27$ & $0.05\pm0.36$  \\ \hline
$0.846$ & \begin{tabular}[c]{@{}c@{}}$^{56}$Fe$(n,n')^{56}$Fe\\ $^{57}$Fe$(n,2n)^{56}$Fe\end{tabular} & $0.856(2^+) \rightarrow $g.s.$(0^+)$                                                                   & $784\pm71$ & $0.17\pm0.01$  & $-0.05\pm0.02$ \\ \hline
$0.931$ & $^{56}$Fe$(n,2n)^{55}$Fe                                                          & $0.931(5/2^-) \rightarrow $g.s.$(3/2^-)$                                                               & $78\pm7$   & $0.20\pm0.02$  & $-0.03\pm0.03$ \\ \hline
$1.038$ & \begin{tabular}[c]{@{}c@{}}$^{56}$Fe$(n,n')^{56}$Fe\\ $^{57}$Fe$(n,2n)^{56}$Fe\end{tabular} & $3.122(4^+) \rightarrow 2.085(4^+)$                                                                  & $72\pm7$   & $0.24\pm0.04$  & $-0.01\pm0.06$ \\ \hline
$1.238$ & \begin{tabular}[c]{@{}c@{}}$^{56}$Fe$(n,n')^{56}$Fe\\ $^{57}$Fe$(n,2n)^{56}$Fe\end{tabular} & $2.085(4^+) \rightarrow 0.846(2^+)$                                                                  & $356\pm32$ & $0.26\pm0.01$  & $-0.06\pm0.01$ \\ \hline
$1.303$ & \begin{tabular}[c]{@{}c@{}}$^{56}$Fe$(n,n')^{56}$Fe\\ $^{57}$Fe$(n,2n)^{56}$Fe\end{tabular} & $3.388(6^+) \rightarrow 2.085(4^+)$                                                                  & $136\pm12$ & $0.31\pm0.02$  & $-0.09\pm0.03$ \\ \hline
$1.408$ & \begin{tabular}[c]{@{}c@{}}$^{54}$Fe$(n,n')^{54}$Fe\\ $^{56}$Fe$(n,2n)^{55}$Fe\end{tabular} & \begin{tabular}[c]{@{}c@{}}$1.408(2^+) \rightarrow $g.s.$(0^+)$\\ $1.408(7/2^-) \rightarrow $g.s.$(3/2^-)$\end{tabular}  & $33\pm3$   & $0.17\pm0.04$  & $-0.08\pm0.05$ \\ \hline
$1.670$ & \begin{tabular}[c]{@{}c@{}}$^{56}$Fe$(n,n')^{56}$Fe\\ $^{57}$Fe$(n,2n)^{56}$Fe\end{tabular} & $3.755(6^+) \rightarrow 2.085(4^+)$                                                                  & $43\pm4$   & $0.42\pm0.04$  & $-0.15\pm0.05$ \\ \hline
$1.810$ & \begin{tabular}[c]{@{}c@{}}$^{56}$Fe$(n,n')^{56}$Fe\\ $^{57}$Fe$(n,2n)^{56}$Fe\end{tabular} & $2.657(2^+) \rightarrow 0.846(2^+)$                                                                  & $51\pm5$   & $0.02\pm0.07$  & $-0.12\pm0.11$  \\ \hline
$2.598$ & \begin{tabular}[c]{@{}c@{}}$^{56}$Fe$(n,n')^{56}$Fe\\ $^{57}$Fe$(n,2n)^{56}$Fe\end{tabular} & $3.445(3^+) \rightarrow 0.846(2^+)$                                                                  & $31\pm3$   & $-0.31\pm0.03$ & $-0.06\pm0.04$ \\ \hline
\end{tabular}
\end{center}
\begin{multicols}{2}

\section{Conclusion}
In this work, differential cross-sections of $\gamma$- emission generated in reactions under the action of $14.1$~MeV neutrons on the nuclei of carbon, aluminum, silicon, calcium, chromium and iron were measured. The measurements were performed using four LaBr$_3$(Ce) scintillation detectors positioned at angles of $25^{\circ}$, $45^{\circ}$, $60^{\circ}$, and $70^{\circ}$ relative to the axis of the generator target -- the center of the sample. A key feature of this study was the implementation of the tagged neutron method. The experiments utilized a neutron generator capable of producing 16 separate tagged neutron beams. Combined with the detector system, this enabled measurements of differential cross sections at $64$ discrete angles in the $17-89^{\circ}$ range. Corrections for multiple neutron scattering and attenuation, $\gamma$-ray attenuation, and total detection efficiency -- calculated using GEANT4 -- were systematically applied. Verification measurements were conducted to validate these correction factors. The analysis yielded angular distribution data for the $4.439$~MeV $\gamma$-line from carbon, $10$ $\gamma$-lines from aluminum reactions, $6$ $\gamma$-lines from silicon reactions, $8$ $\gamma$-lines from calcium reactions, $16$ $\gamma$-lines from titanium reactions, $6$ $\gamma$-lines from chromium reactions, and $14$ $\gamma$-lines from iron reactions. All angular distributions were approximated through expansion in even-order Legendre polynomials, followed by full solid-angle integration to determine total emission cross sections. The total systematic uncertainty of the obtained data was estimated to be $9$\,\%.

\section{Acknowledgements}
We thank D.~N.~Borisov and S.~I.~Negovelov for their help in preparing the experiments.

\end{multicols}

\vspace{15mm}

\vspace{-1mm}
\centerline{\rule{80mm}{0.1pt}}
\vspace{2mm}

\begin{multicols}{2}
%\bibliography{mybibfile}

\begin{thebibliography}{90}

\vspace{3mm}

\bibitem{Karolina2022} Karolina Kolos, Vladimir Sobes, Ramona Vogt et al, Current nuclear data needs for applications, Phys. Rev. Research 4 (2022) 021001; https://doi.org/10.1103/PhysRevResearch.4.021001       

\bibitem{Ruskov2017} I. Ruskov, Yu. Kopatch, V. Bystritsky, V. Skoy, V. Shvetsov, F.-J. Hambsch, S. Oberstedt, R. Capote Noy, D. Grozdanov, TANGRA collaboration, Tangra - an experimental setup for basic and applied nuclear research by means of 14.1 MeV neutrons, ND2016, EPJ Web Conf. 146 (2017) 03024, https://doi.org/10.1051/epjconf/201714603024      

\bibitem{Ruskov2021} I. Ruskov, Yu. Kopach, V. Bystritsky, V. Skoy, D. Grozdanov, N. Fedorov, T. Tretyakova, F. Aliev, C. Hramco, V. Slepnev, N. Zamyatin, A. Gandhi, D. Wang, A. Kumar, E. Zubarev, E. Bogolubov, Y. Barmakov, TANGRA collaboration, TANGRA multidetector systems for investigation of neutron-nuclear reactions at the JINR Frank Laboratory of Neutron Physics, EPJ Web Conf. 256 (2021) 00014, https://doi.org/10.1051/epjconf/202125600014      

\bibitem{Valkovic2016c2} Vladivoj Valkovic (Ed.), 14 MeV Neutrons: Physics and Application, CRC Press, Boca Raton, Chap. 2. (2016) 17-33; https://doi.org/10.1201/b18837       

\bibitem{Valkovic2016c7} Vladivoj Valkovic (Ed.), 14 MeV Neutrons: Physics and Application, CRC Press, Boca Raton, Chap. 7 (2016) 256-392; https://doi.org/10.1201/b18837       

\bibitem{Bolshakov2020} Ilya Bolshakov, Maxim Kolesnik, Maxim Sorokin, Vladislav Kremenets, Egor Razinkov, et al. Application of Tagged Neutron Method for Element Analysis of Phosphorus Ore. Int. J. Miner. Process. Extr. Metall. 5(4) (2020) 54-59. doi: https://doi.org/10.11648/j.ijmpem.20200504.11

\bibitem{Alexakhin2022} V.Y. Alexakhin, A.I. Akhunova, E.A. Razinkov, et al. Application of the Tagged Neutrons Method for the Analysis of Material on a Conveyor., Phys. Atom. Nuclei 85 (2022) 1866; https://doi.org/10.1134/S1063778822100039       

\bibitem{Bishnoi2020} Bishnoi, S., Patel, T., Thomas, R.G. et al. Study of tagged neutron method with laboratory D-T neutron generator for explosive detection. Eur. Phys. J. Plus 135 (2020) 428; https://doi.org/10.1140/epjp/s13360-020-00402-y

\bibitem{Grozdanov2018} D.N. Grozdanov, N.A. Fedorov, V.M. Bystritski et al, Measurement of Angular Distributions of Gamma Rays from the Inelastic Scattering of 14.1-MeV Neutrons by Carbon and Oxygen Nuclei, Physics of Atomic Nuclei 81 (2018) 588; https://doi.org/10.1134/S106377881805006X

\bibitem{Fedorov2019} N.A. Fedorov, T.Y. Tretyakova, V.M. Bystritsky et al, Investigation of Inelastic Neutron Scattering on $^{27}$Al Nuclei, Phys. Atom. Nuclei 82 (2019) 343; https://doi.org/10.1134/S1063778819040094

\bibitem{Fedorov2021} N.A. Fedorov, D.N. Grozdanov, Y.N. Kopatch, et al., Inelastic scattering of 14.1 MeV neutrons on iron, Eur. Phys. J. A 57 (2021) 194. https://doi.org/10.1140/epja/s10050-021-00503-x

\bibitem{Dashkov2022} I.D. Dashkov, N.A. Fedorov, D.N. Grozdanov et al., Measurement of the Angular Distribution of 14.1 MeV Neutrons Scattered by Carbon Nuclei, Bull. Russ. Acad. Sci. Phys. 86, 893-897 (2022). https://doi.org/10.3103/S1062873822080056

\bibitem{Kopatch2023} Kopach, Y.N., Sapozhnikov, M.G. Applications of the Tagged Neutron Method for Fundamental and Applied Research. Phys. Part. Nuclei 55, 55--102 (2024). https://doi.org/10.1134/S106377962401009X

\bibitem{Kopatch2025} Kopatch, Y.N., Grozdanov, D.N., Fedorov, N.A. et al. Measurement of the $\gamma$-Quanta Emission Cross Sections in Reactions $(n, X\gamma)$ for $^{28}$Si and $^{16}$O Using the Method of Tagged Neutrons with $E_n$ = $14.1$~MeV. Phys. Part. Nuclei Lett. 22, 276-279 (2025). https://doi.org/10.1134/S154747712470225X

\bibitem{Scherrer1953} V. E. Scherrer, R. B. Theus, W. R. Faust, Gamma Rays from Interaction of 14-Mev Neutrons with Various Materials, Phys. Rev. 91 (1953) 1476; https://doi.org/10.1103/PhysRev.91.1476

\bibitem{Benveniste1960} J. Benveniste, A.C. Mitchell, C.D. Schrader, J.H. Zenger, Gamma rays from the interaction of 14-MeV neutrons with carbon, Nucl. Phys. 19 (1960) 448; https://doi.org/10.1016/0029-5582(60)90255-8

\bibitem{Stewart1964} D.T. Stewart, P.W. Martin, Gamma rays from the interaction of 14 MeV neutrons with $^{12}$C and $^{24}$Mg, Nucl. Phys. 60 (2) (1964) 349; https://doi.org/10.1016/0029-5582(64)90669-8

\bibitem{Morgan1964} I.L. Morgan, J.B. Ashe, D.O. Nellis, Angular distribution of gamma rays from C, O, and N at En=14.8 MeV, Div. of Tech. Info. U.S. AEC Reports, No. 22012, p.158 (1964), Texas Nuclear Corp., Austin, TX, USA, EXFOR data: https://www-nds.iaea.org/EXFOR/12695

\bibitem{Engesser1967} F. C. Engesser and W. E. Thompson, Gamma rays resulting from interactions of 14.7 MeV neutrons with various elements, J. Nucl. Energy, 21 (6) (1967) 487;   https://doi.org/10.1016/0022-3107(67)90020-2

\bibitem{Clayeux1969} G. Clayeux,G. Grenier, Recoil spectra produced by 14.1 MeV neutrons, Centre d'Etudes de Limeil, Villeneuve-Saint-Georges, France, Centre d`Etudes Nucleaires, Saclay Reports, No.3807 (1969)

\bibitem{Martin1971} T.C. Martin, G.H. Williams, Production cross section for the 4.43 gamma-ray from C, Oak Ridge Operations Office, contract report, No. 2791-32, p.7 (1971), Texas Nuclear Corp., Austin, TX, USA

\bibitem{Spaargaren1971} D. Spaargaren, C.C. Jonker, Angular correlations in inelastic neutron scattering by carbon at 15.0 MeV, Nucl. Phys. A 161(2) (1971) 354; https://doi.org/10.1016/0375-9474(71)90374-5

\bibitem{Rogers1975} V.C. Rogers, V.J. Orphan, C.G. Hoot, V.V. Verbinski, Gamma-Ray Production Cross Sections for Carbon and Nitrogen from Threshold to 20.7 MeV, Nucl. Sci. Eng. 58 (3) (1975) 298; https://doi.org/10.13182/NSE75-A26779

\bibitem{Bezotosnyi1975} V. M. Bezotosnyi, V. M. Gorbachev, L. M. Suvorov, M.S. Shvetsov, The cross section of gamma-rays production at inelastic interaction of the 14 MeV neutrons with different nuclei, Yadernye Konstanty, 19 (1975) 77.

\bibitem{Dickens1977} J.K. Dickens, G.L. Morgan, G.T. Chapman, T.A. Love, E. Newman, F.G. Perey, Cross Sections for Gamma-Ray Production by Fast Neutrons for 22 Elements Between Z = 3 and Z = 82, Nucl. Sci. Eng. 62 (1977) 515; https://doi.org/10.13182/NSE77-A26989

\bibitem{Zong1979} S. Zong-Ren et al., Gamma-Ray Production Cross Sections from Interactions of 14.9 MeV Neutrons with C, F, Al, Si, Fe and Cu, Chin. J. of Nucl. Phys., 1 (1979)  45.

\bibitem{Hino1979} Y. Hino, S. Itagaki, T. Yamamoto, K. Sugiyama, Cross section for the reaction of $^9$Be$(n,t\gamma)$ and $^{12}$C$(n,n\gamma)$ between 13 and 15 MeV, Tohoku Univ., Sendai, Japan, Tohoku Univ., Dept. of Nucl. Engineering Reports, No. 34 (1979).

\bibitem{Murata1988} I. Murata, J. Yamamoto, A. Takahashi, Differential cross sections for gamma-ray production by 14 MeV neutrons with several elements in structural materials, Conf. on Nucl. Data For Sci. and Technol., Mito 1988, p. 275 (1988), Japan. https://wwwndc.jaea.go.jp/nd1988/Mito\%20Conf/300/12916-0275.pdf

\bibitem{Zhou1989} Zhou Hongyu, Yan Yiming, Tanglin, Wen Shenlin et al, Production cross section measurement of discrete gamma-ray at 90 deg. for interactions of 14.9 MeV neutrons with carbon and niobium, Chinese J. of Nuclear Physics 11(2) (1989) 63.

\bibitem{Hasegawa1991} K. Hasegawa, M. Mizumoto, S. Chiba, M. Igashira, and H. Kitazawa, Gamma-ray production cross section measurements of some structural materials between 7.8 and 13.0 MeV, in Proceedings of the Conference on Nuclear Data for Science and Technology, Juelich, 1991, p. 329

\bibitem{Lashuk1994} A.I. Lashuk, I.P. Sadokhin, Gamma-quanta production cross-sections at inelastic scattering of the neutrons on the nuclei of reactor construction materials, Vop. At. Nauki i Tekhn., Ser.Yaderno-Reak. Konstanty, 1 (1994) 26.

\bibitem{Kadenko2016} I.M. Kadenko, V.A. Plujko, B.M. Bondar, O.M. Gorbachenko, B.Yu. Leshchenko, K.M. Solodovnyk, Gamma-rays from $^{nat}$Sn and $^{nat}$C induced by fast neutrons, Yaderna Fizika ta Energetika, 17 (2016) 349

\bibitem{McEvoy2021} A. M. McEvoy, H. W. Herrmann, Y. Kim et al, $^{13}$C$(n,2n\gamma)^{12}$C $\gamma$-ray production in the 14--16 MeV incident neutron energy range, Phys. Rev. C 103 (2021) 064607; https://doi.org/10.1103/PhysRevC.103.064607

\bibitem{Kelly2021} K. J. Kelly, M. Devlin, J. M. O'Donnell, E. A. Bennett, Correlated $n-\gamma$ angular distributions from the $Q$=4.4398 MeV $^{12}$C$(n,n'\gamma)$ reaction for incident neutron energies from 6.5 MeV to 16.5 MeV, Phys. Rev. C 104 (2021) 064614; https://doi.org/10.1103/PhysRevC.104.064614

\bibitem{Kelly2023} K. J. Kelly, M. Devlin, J. M. O'Donnell, E. A. Bennett, M. Paris, and P. A. Copp, Measurement of the cross section of the $Q$=4.4398 MeV $^{12}$C$(n,n'\gamma)$ reaction from threshold to 16.5 MeV using $\gamma$ and correlated $n-\gamma$ detection, Phys. Rev. C 108 (2023) 014603; https://doi.org/10.1103/PhysRevC.108.014603

\bibitem{Gordon2025} J. M. Gordon, B. L. Goldblum, J. A. Brown et al, $^{12}C(n,n'_1\gamma)$ partial $\gamma$-ray cross section measured using the GENESIS array, Phys. Rev. C 111 (2025) 044608; https://doi.org/10.1103/PhysRevC.111.044608

\bibitem{Morgan1977} G.L.Morgan, T.A.Love, J.K.Dickens, F.G.Perey, Cross Sections for Gamma-Ray Production by Fast Neutrons for 22 Elements Between Z = 3 and Z = 82, Nucl. Sci. Engineering, 62 (1977) 515; http://dx.doi.org/10.13182/NSE77-A2698

\bibitem{HINO1978} HINO Y., YAMAMOTO T., SAITO T., ARAI Y., ITAGAKI S., \& SUGIYAMA K.,  Gamma-Ray Production Cross Sections for Aluminum and Copper at 5.3-MeV Neutron Energy. Journal of Nuclear Science and Technology, 1978, 15(2), 85-90. https://doi.org/10.1080/18811248.1978.9735494

\bibitem{Grozdanov2026} D.N.Grozdanov et al., Measurement of differential and total cross sections for
scattering of 14.1 MeV neutrons on 12C nuclei, EPJ A, 2026, to be published

\bibitem{Burymov1969} E.M.Burymov, Cross-section for excitation of $^{16}$O, $^{27}$Al, $^{52}$Cr and $^{64,66,68}$Zn levels by $14.6$~MeV neutrons, Soviet Journal of Nuclear Physics 9 (1969) 546

\bibitem{Bochkarev1965} V.N. Bochkarev, V.V. Nefedov, Gamma radiation from the inelastic interaction of 14-MeV neutrons with magnesium, aluminum, and sulfur nuclei, Soviet Journal of Nuclear Physics 1 (1965) 574.

\bibitem{Pavlik1998} A. Pavlik, H. Hitzenberger-Schauer, H. Vonach, M. B. Chadwick, R. C. Haight, R. O. Nelson, and P. G. Young, $^{27}$Al$(n,x\gamma)$ reactions for neutron energies from 3 to 400 MeV, Phys. Rev. C  57 (1998) 2416; https://doi.org/10.1103/PhysRevC.57.2416

\bibitem{Hlavac1999} S. Hlavac, M. Benovic, E. Betak, L. Dostal, I. Turzo, S.P. Simakov, Cross Sections for Discrete $\gamma$ Ray Production in Interactions of 14.6 MeV Neutrons with Light and Medium Heavy Nuclei, IAEA Nucl.Data Section report to the I.N.D.C., No.412, p.12 (1999), Austria; http://www-nds.iaea.org/publications/indc/indc-nds-0412/

\bibitem{Hoot1975} V. C. Rogers, V. J. Orphan, C. G. Hoot, V. V. Verbinski, D. G. Costello, and S. J. Friesenhahn,Spectral gamma-ray production cross-section measurements from threshold to 20 MeV, Conf. on Nucl.Cross-Sect.and Techn.,Washington 1975, Vol.2, p.766 (1975), USA; https://inis.iaea.org/records/gcqdx-arb55

\bibitem{Zhou1997} Zhou, H., \& Huang, G. (1997). Study of Total Discrete Gamma Radiation from Aluminum under 14.9-MeV Neutron Bombardment. Nuclear Science and Engineering, 125(1), 61-74. https://doi.org/10.13182/NSE97-A242545

\bibitem{Nyberg1971} K Nyberg-Ponnert, B Jonsson, and I Bergqvist, Gamma Rays Produced by the Interaction of 15 MeV Neutrons in N, O, Mg and Al, 1971, Phys. Scr., 4, 165, https://dx.doi.org/10.1088/0031-8949/4/4-5/004

\bibitem{Hongyu1986} Z. Hongyu et al., "Gamma ray production cross sections for the interactions of 14.9 MeV neutrons with C,Al,V,Fe and Nb at 90 degrees," Institute of Low Energy Nuclear Physics, Austria, Chinese report to the I.N.D.C. 10, 1986

\bibitem{Prud'homme1960} J. T. Prud'homme, I. L. Morgan, J. H. MC. Crary, and J. B. Ashe, "A study of neutrons and gamma rays from neutron induced reactions in several elements," Texas Nuclear Corp., Austin, TX, United States of America, USA, Air Force Spec. Weap. Center Kirtland A.F.B. Repts. 60-30, 1960.

\bibitem{Boromiza2020} M. Boromiza, C. Borcea, P. Dessagne, D. Ghita, T. Glodariu, G. Henning, M. Kerveno, N. Marginean, C. Mihai, Nucleon inelastic scattering cross sections on $^{16}$O and $^{28}$Si, Phys. Rev. C 101 (2020) 024604; https://doi.org/10.1103/PhysRevC.101.024604

\bibitem{Martin1965} P.W. Martin, D.T. Stewart, Gamma-ray yields from inelastic neutron scattering of 14.1 MeV neutrons from sodium, magnesium, silicon, sulphur, manganese and iron, J. Nucl. Energy 19 (6) (1965) 447; https://doi.org/10.1016/0368-3230(65)90053-4

\bibitem{Abbondanno1973} U. Abbondanno, R. Giacomich, M. Lagonegro, and G. Pauli, "Gamma rays resulting from nonelastic processes of 14.2 MeV neutrons with sodium, magnesium, silicon, sulphur, titanium, chromium and iron," J. Nucl. Energy, 27 (4) 1973 4, https://doi.org/10.1016/0022-3107(73)90058-0

\bibitem{Connell1975} K.A. Connell, A.J. Cox, The use of a small accelerator to study the gamma rays associated with the inelastic scattering of 14 MeV Neutrons in $^{28}$Si, $^{32}$S and $^{48}$Ti, The International Journal of Applied Radiation and Isotopes 26 (2) (1975) 71; https://doi.org/10.1016/0020-708X(75)90105-2

\bibitem{Bezotosnyj1980} V.M. Bezotosnyj, V.M.Gorbachjov, M.S.Shvetsov, L.M.Surov, The spectra and formation cross-sections of the discrete gamma-lines in the nonelastic interaction of 14 MeV neutrons with Mg, Si, P, S, Ti and Zn nuclei, Nat. Issl. Tsentr "Kurchatovskii Institut", Moskva, Russia, USSR report to the I.N.D.C., No.169, Vol.(2), p.21 (1980)

\bibitem{Drosg2002} M. Drosg, R. C. Haight, and D. M. Drake, "Double-differential gamma-ray production: cross sections and spectra of Al, Si and Fe for 8.51, 10.00, 12.24 and 14.24 MeV neutrons," Los Alamos National Laboratory, NM, United States of America, USA, Los Alamos Scientific Lab. Reports 02-16, 2002

\bibitem{Zhou2011} Hong-Yu Zhou, Fu-Guo Deng, Wei Cheng, Feng-Shou Zhang, Qiang Zhao, Jun Su, Li-Ming Dong, Qing Zhu, Guo-Ying Fan, Associated gamma radiation in interaction of 14.9 MeV neutrons with natural silicon, Nucl. Instrum. Methods Phys. Res. A 648 (1) (2011) 192; https://doi.org/10.1016/j.nima.2011.04.014

\bibitem{Negret2013} A. Negret, C. Borcea, D. Bucurescu et al, Cross sections for inelastic scattering of neutrons on $^{28}$Si and comparison with the $^{25}$Mg$(\alpha,n)^{28}$Si reaction, Phys. Rev. C 88 (2013) 034604; https://doi.org/10.1103/PhysRevC.88.034604

\bibitem{Guoying1992} Guoying Fan, Hongyu Zhou, Xiaoge Zhu, Ming Hua, Shenlin Wen, Liqiao Lan, Shuzhen Liu,  Progress in the Measurement of Gamma-Ray Production Cross Sections Induced by 14.9 MeV Neutrons. In: Qaim, S.M. (eds) Nuclear Data for Science and Technology. Research Reports in Physics. Springer, Berlin, Heidelberg. (1992). https://doi.org/10.1007/978-3-642-58113-7\_95

\bibitem{Arya1967} A. P. Arya, D. L. Campbell, and R. D. Wilson, Gamma yield from 14.3-MeV neutron inelastic scattering and comparison with quadrupole transitions, Bull. Am. Phys. Soc.,  12 (1967) HD4.

\bibitem{Breunlich1971} W. Breunlich, G.Stengl, H.Vonach, Determination of $(n,n'\gamma)$ cross sections for $14.4$~MeV neutrons in the mass range A=46 to 88, Zeitschrift fuer Naturforschung, Section A 26 (1971) 451; http://dx.doi.org/10.1515/zna-1971-0315

\bibitem{Morgan1978} G. L. Morgan, D. C. Larson, The Ti($n,x\gamma$) reaction cross section for incident neutron energies between 0.3 and 20.0 MeV, Oak Ridge National Laboratory, Oak Ridge, TN, USA, Oak Ridge National Lab. technical memo, No.6323 (1978).https://inis.iaea.org/records/806kn-26t18

\bibitem{Dashdorj2005} D. Dashdorj, P. E. Garret, J. A. Becker et al, $^{48}$Ti$(n,xnypz\alpha\gamma)$ Reactions for Neutron Energies up to 250 MeV, AIP Conf. Proc. 769 (2005) 1035; https://doi.org/10.1063/1.1945183

\bibitem{Dashdorj2007} D. Dashdorj, G.E. Mitchell,  J. A. Becker et al, Gamma-Ray Production Cross Sections in Multiple Channels for Neutron-Induced Reaction on 48Ti for En = 1 to 200 MeV, Nucl. Sci. Eng. 157(1) (2007) 65; https://doi.org/10.13182/NSE07-A2713

\bibitem{Olacel2017} A. Olacel, F. Belloni, C. Borcea et al, Neutron inelastic scattering measurements on the stable isotopes of titanium, Phys. Rev. C 96 (2017) 014621; https://doi.org/10.1103/PhysRevC.96.014621

\bibitem{Kinney1972} H. O. M. Kinney, "Inelastic scattering of 14.4 MeV Neutrons from several elements," Thesis: Mc Kinney, University of West Virginia, Morgentown, WV, United States of America, 1972.

\bibitem{Grenier1974} G. Grenier, B. Duchemin, D. Parisot, Differential gamma-ray production cross sections measured in Si$(n,X\gamma)$, Cr$(n,X\gamma)$ and Ni$(n,X\gamma)$ reactions for incident neutrons between 3 and 7 MeV and also at 14.1 MeV (in French), CEA/DAM Ile-de-France, Bruyeres-le-Chatel, Arpajon, France, Report from CEC-Countries and CEC to NEANDC, No.161U, 1974. https://inis.iaea.org/records/jwywe-zwf06

\bibitem{Voss1975} F.Voss, S.Cierjacks, D.Erbe, G.Scmalz, Measurement of the gamma-ray production cross section from inelastic neutron scattering in some chromium and nickel isotopes between 0.5 and 10 MeV, Conf. on Nucl. Cross-Sect. and Techn., Washington 1975, p.278 (1975), USA.

\bibitem{Yamamoto1978} T. Yamamoto, Y. Hino, S. Itagaki, and K. Sugiyama, "Gamma-Ray Production Cross Sections for Interactions of 14.8 MeV Neutrons with O, Na, Al, Cl, Cr, Fe, Ni, Cu and Pb," J. Nucl. Sci. Technol., vol. 15, no. 11, Art. no. 11, 1978, doi: 10.1080/18811248.1978.9735594

\bibitem{Oblozinsky1992} P. Oblozinsky, S. Hlavac, G. Maino, A. Mengoni, Gamma-ray production from $^{52}$Cr$(n,x\gamma)$ reactions at 14.6 MeV, Il Nuovo Cimento A 105 (1992) 965; https://doi.org/10.1007/BF02730838

\bibitem{Mihailescu2007} L.C. Mihailescu, C. Borcea, A.J. Koning, A.J.M. Plompen, High resolution measurement of neutron inelastic scattering and (n,2n) cross-sections for $^{52}$Cr, Nucl. Phys. A 786 (1-4) (2007) 1; https://doi.org/10.1016/j.nuclphysa.2007.01.004

\bibitem{Bostrom1959} M. A. Bostrom, I. L. Morgan, J. T. Prud'homme, and P. L. Okhuysen, "Interaction of fast neutrons in iron, lead, oxygen and hydrogen," Texas Nuclear Corp., Austin, TX, United States of America, USA, Wright Air Devel. Centre Reports 50-31, 1959

\bibitem{Joensson1969} B. Joensson, K. Nyberg, and I. Bergquist, "High Resolution Measurements of Gamma Rays Produced by 15 MeV Neutrons," Ark. Foer Fys., vol. 39, p. 295, 1969

\bibitem{Sukhanov1970} B. I. Sukhanov and N. P. Tkach, "Gamma rays in inelastic interaction of 14.-MeV            neutrons with N,O,Al, and Fe nuclei," Sov. J. Nucl. Phys., vol. 11, p. 17, 1970.

\bibitem{Degtyarev1977} A. P. Degtyarev, V. V. Kravtsov, and . A. Prokopets, "Gamma-rays connected with the nonelastic processes of the deuterium-tritium neutrons on the nuclei Na, Fe", Kyivsky Natsionalny Univ. "Taras Shevchenko", Kyiv, Ukraine, USSR report to the I.N.D.C. INDC(CCP)-118, 1977

\bibitem{Drake1978} D. M. Drake, E. D. Arthur, and M. G. Silbert, "Cross Sections for Gamma-Ray Production by 14-MeV Neutrons," Nucl. Sci. Eng., vol. 65, no. 1, Art. no. 1, 1978, https://doi.org/10.13182/NSE78-A27125

\bibitem{Xiamin1982} S. Xiamin, S. Ronglin, X. Jinqiang, W. Yongshun, and D. Dazhao, "Measurements of the induced gamma ray cross-sections by 14.2 MeV neutrons with Fe, Ni and Cu," in Proceedings of Conference on Nuclear Data for Science and Technology, Antwerpen, Belgium, 1982, p. 373

\bibitem{Shalabi1983} B. Al-Shalabi, A. J. Cox, The gamma rays associated with the inelastic scattering of 14 MeV neutrons in large samples of iron, Nucl. Instrum. Methods Phys. Res. 205 (3) (1983) 495, https://doi.org/10.1016/0167-5087(83)90015-7

\bibitem{Hlavac1983} S. Hlavac and P. Oblozinsky, "Gamma-ray production cross sections and gamma-ray multiplicities from Fe and Ni bombarded with 14.6 MeV neutrons," Slovak Academy of Sciences, Physics Inst., Bratislava, Slovakia, Austria, CSSR report to the I.N.D.C. 5, 1983

\bibitem{Jinqiang1988} X. Jinqiang, C. Zhong, W. Yongshun, and C. Qun, Measurement of Differential Production Cross Sections for Discrete Gamma Rays from Fe$(n,x\gamma)$ Reaction, Chin. J. of Nucl. Phys. 10(3) (1988) 284

\bibitem{Sakane1999} H. Sakane, Y. Kasugai, F. Maekawa, Y. Ikeda, and K. Kawade, "Measurement of discrete gamma-ray production cross sections for interaction of 14-MeV neutrons with Mg, Al, Si, Ti, Fe, Ni, Cu, Nb, Mo and Ta," in JAERI Conference proceedings, Japan, 1999, p. 216

\bibitem{Negret2014} A. Negret et al., Cross-section measurements for the $^{56}$Fe$(n,xn\gamma)$ reactions, Phys. Rev. C 90, 034602 (2014) https://doi.org/10.1103/PhysRevC.90.034602

\bibitem{Broder1970} D. L. Broder, A. F. Gamaliy, A. I. Lashuk, I. P. Sadokhin, Inelastic Neutron Scattering -- $(n,n'\gamma)$- by Fluorine, Iron, Cobalt, Nickel and Tantalum, Nuclear Data for Reactors -- 1970, Proceedings Series - International Atomic Energy Agency, IAEA, Vienna (1970) https://www-nds.iaea.org/publications/proceedings/sti-pub-259-Vol2.pdf

\bibitem{Western1965} G. T. Western, R. C. Baird,F. L. Gibbons,  J. R. Williams, The interactions of 14.6-MeV neutrons in iron, Lithium-7, and carbon, General Dynamics, Fort Worth, Techn. Reports, No.64-140 (1965) https://apps.dtic.mil/sti/html/tr/AD0614451/index.html

\bibitem{Mitsuda2002} Mitsuda M., Kondo T., Murata I., Takahashi A., Measurements of Secondary Gamma-Ray Production Cross Sections for $^{nat}$Fe, $^{51}$V, $^{nat}$Mo, $^{nat}$Zr, $^{nat}$Ni and $^{181}$Ta with Hp-Ge Detector Induced by DT Neutrons. Journal of Nuclear Science and Technology, 39(2) (2002) 437-440. https://doi.org/10.1080/00223131.2002.10875134

\bibitem{Antalik1980} R. Antalik, S. Hlavac, P. Oblozinsky, A study of $^{56}$Fe$(n,xn\gamma)$ reactions at $14.6$~MeV, vol. 6. VEDA, Publishing House of the Slovak Academy of Sciences., 1980, pp. 277-287. https://inis.iaea.org/records/btnad-90472

\bibitem{Nelson2005} R. O. Nelson, N. Fotiades, M. Devlin, J. A. Becker, P. E. Garrett, W. Younes, Cross-Section Standards for Neutron-Induced Gamma-Ray Production in the MeV Energy Range, AIP Conf. Proc. 769(1) (2005) 838-841. https://doi.org/10.1063/1.1945136

\bibitem{Simakov1998} S. P. Simakov, A. Pavlik, H. Vonach, S. Hlavac, Status of experimental and evaluated discrete $\gamma$-rays production at En=14.5 MeV, final report of research contract 7809/RB, performed under the CRP on measurement, Calculation and Evaluation of Photon Production Data, INDC(CCP)- 413 (1998); https://inis.iaea.org/records/c44gw-yfn27

\bibitem{Bogolyubov2017} E. P. Bogolyubov, A. V. Gavryuchenkov, M. D. Karetnikov, D. I. Yurkov, V. I. Ryzhkov, V. I. Zverev, Neutron generators and DAQ system for tagged neuron technology. Proceedings of the XXVI International Symposium on Nuclear Electronics \& Computing (NEC'2017), Becici, Budva, Montenegro, September 25 - 29, 2017, pp. 176-181. https://ceur-ws.org/Vol-2023/176-181-paper-27.pdf

\bibitem{Prusachenko2025} P. S. Prusachenko, D. N. Grozdanov, N. A. Fedorov et al., Characterization of an EJ-200 plastic scintillator array for experiments with 14-MeV tagged neutrons using the carbon and polyethylene samples, Nucl. Instrum. Methods Phys. Res. A 1072 (2025) 170143; https://doi.org/10.1016/j.nima.2024.170143

\bibitem{Capote2009} R. Capote, M. Herman, P. Oblozinsky et. al, RIPL - Reference Input Parameter Library for Calculation of Nuclear Reactions and Nuclear Data Evaluations,  NDS, 110 (12), (2009) 3107.  https://doi.org/10.1016/j.nds.2009.10.004

\bibitem{ENSDF} Evaluated Nuclear Structure Data Files (ENSDF) database; https://doi.org/10.18139/nndc.ensdf/1845010

\bibitem{Agostinelli2003} S. Agostinelli, et al., GEANT4 Collaboration, Geant4--a simulation toolkit, Nucl. Instrum. Methods Phys. Res. 506(3) (2003) 250, https://doi.org/10.1016/S0168-9002(03)01368-8

\bibitem{Allison2016} J. Allison, e. al., GEANT4 collaboration, recent developments in Geant4, Nucl. Instrum. Methods Phys. Res. 835 (2016) 186, https://doi.org/10.1016/j.nima.2016.06.125

\bibitem{Richard2003} Richard M. Lindstrom, Richard B. Firestone and R. Paviotti-Corcuera, Development of a Database for Prompt $\gamma$-ray Neutron Activation Analysis, Summary Report of the Third Research Coordination Meeting (INDC(NDS)-443), IAEA Nuclear Data Section, Vienna, Austria, 2003; https://www-nds.iaea.org/pgaa/Annex1/INDC\_NDS\_443.pdf

\bibitem{Koning2023} A. Koning, S. Hilaire, \& S. Goriely, TALYS: modeling of nuclear reactions, Eur. Phys. J. A 59, 131 (2023); https://doi.org/10.1140/epja/s10050-023-01034-3

\bibitem{Brown2018} D.A. Brown et al., ENDF/B-VIII.0: The 8th Major Release of the Nuclear Reaction Data Library with CIELO-project Cross Sections, New Standards and Thermal Scattering Data, Nucl. Data Sheets 148 (2018) 1; https://doi.org/10.1016/j.nds.2018.02.001

\bibitem{Iwamoto2023} O. Iwamoto, N. Iwamoto, S. Kunieda, F. Minato, S. Nakayama, Y. Abe, et al., Japanese evaluated nuclear data library version 5: JENDL-5, J. Nucl. Sci. Technol. 60(1) (2023) 1; https://doi.org/10.1080/00223131.2022.2141903


\end{thebibliography}

\end{multicols}

\clearpage
\end{CJK*}
\end{document}